\documentclass[preprint]{elsarticle}
\usepackage[utf8]{inputenc}
\usepackage{subfigure,verbatim}
\usepackage{amsmath,bm,amssymb}
\usepackage{natbib}

\usepackage[dvipsnames]{xcolor}

\newcommand{\moritz}[1]{{\textcolor{black}{#1}}}

\title{Effects of anisotropy on the geometry 
of tracer particle trajectories 
in turbulent flows}

\author[inst1]{Yasmin Hengster \corref{cor1}}
\cortext[cor1]{Corresponding authors}
\author[2]{Martin Lellep } 
\author[3]{Julian Weigel}
\author[4,5]{Matthew Bross}
\author[6]{Johannes Bosbach}
\author[6]{Daniel Schanz}
\author[6,7]{Andreas Schr{\"o}der}
\author[6]{Florian Huhn}
\author[6]{Matteo Novara}
\author[8]{Daniel Garaboa Paz}
\author[5]{Christian J. Kähler} 
\author[inst1]{ \linebreak Moritz Linkmann \corref{cor1}}

\affiliation[inst1]{organization={School of Mathematics and Maxwell Institute for Mathematical Science, University of Edinburgh},
            addressline={Mayfield Rd},
            city={Edinburgh},
            postcode={EH9 3FD},
            country={United Kingdom}}
            
\affiliation[2]{organization={SUPA, School of Physics and Astronomy, University of Edinburgh},
addressline={Peter Guthrie Tait Rd},
postcode={EH9 3FD},
city={Edinburgh},
country={United Kingdom}}

\affiliation[3]{organization={Department for Physics and Astronomy, University of Heidelberg},
postcode={D-69120},
city={Heidelberg},
country={Germany}}

\affiliation[4]{organization={Fluids Research Department, Applied Research Laboratory (ARL), Pennsylvania State University, State College PA }, country={United States}}

\affiliation[5]{organization={Institute of Fluid Mechanics and Aerodynamics, Universität der Bundeswehr München}, city={Neubiberg},country={Germany}}

\affiliation[6]{organization={German Aerospace Center (DLR), Institute of Aerodynamics and Flow Technology}, city={Göttingen}, country={Germany}}

\affiliation[7]{organization={Brandenburgische Technische Universität (BTU)}, city={Cottbus-Senftenberg}, country={Germany}}

\affiliation[8]{organization={Group of Non-linear Physics, University of Santiago de Compostela}, country={Spain}}

\begin{document}

\begin{abstract} Using curvature and torsion to describe Lagrangian
	trajectories gives a full description of these as well as an insight
	into small and large time scales as temporal derivatives up to order 3
	are involved. One might expect that the statistics of these properties
	depend on the geometry of the flow. Therefore, we calculated curvature
	and torsion probability density functions (PDFs) of experimental
	Lagrangian trajectories processed using the Shake-the-Box algorithm of
	turbulent von Kármán flow, Rayleigh-Bénard convection and a
	zero-pressure-gradient turbulent boundary layer over a flat plate.  The results
	for the von Kármán flow compare well with previous experimental results for the
	curvature PDF and numerical simulation of homogeneous and isotropic
	turbulence for the torsion PDF. Results for  Rayleigh-Bénard convection  
	agree with those obtained for Kármán flow, while results
	for the logarithmic layer within the boundary
	layer differ slightly, and we provide a potential explanation.
        To detect and quantify the effect of anisotropy either resulting from a mean flow or large-scale coherent motions on the geometry or tracer particle trajectories, we introduce the curvature vector. We connect its statistics with those of velocity fluctuations and demonstrate 
        that strong large-scale motion in a given spatial direction results in meandering rather than helical trajectories. 
\end{abstract}

\maketitle

\section{Introduction}

An important feature of turbulent flows in nature and engineering applications is large-scale spatio-temporal coherence, more precisely, the presence of persistent large-scale structures. Such {\em turbulent superstructures} occur in turbulent boundary layers in the laboratory \cite{Kim1999,Adrian2000,Guala2006,Hutchins2007a} or the atmosphere \cite{Hutchins2012}, 
and also in Rayleigh-B\'enard convection \cite{Pandey2018,Stevens2018,Schneide2018,Krug2020},
for instance. They influence mixing and extreme events and, at least in case of turbulent boundary layers, contribute significantly to momentum exchange and kinetic energy of the flow \cite{Monty2007,Hutchins2011} 
and to the Reynolds stress \cite{Ganapathisubramani2003,Guala2006}.
One particular question that is of interest from a fundamental and from a turbulence modelling perspective is the connection between turbulent superstructures and extreme small-scale velocity fluctuations \cite{Hutchins2007b,Bross2019}, in particular near a solid boundary. 
Unusually high values of torsion and curvature in tracer particle trajectories indicate the presence
of intense small-scale vortices. These can usually not be resolved
in numerical simulations at parameters relevant in industrial applications or atmospheric physics, that is, their effect needs to be incorporated into turbulence models.
There is no doubt that large-scale spatio-temporally coherent structures influence the motion of Lagrangian particle trajectories, and presumably the statistics of instantaneous curvature and torsion thereof. Curvature is a multi-scale observable in the sense that connects velocity and acceleration, hence its statistics may contain information regarding the effect of large-scale dynamics on the small scales.  

\moritz{
The purpose of this paper is to compare curvature and torsion statistics across different turbulent flows, including for the first time a turbulent boundary layer, to provide (a) baseline statistics for more refined analyses in connection with large-scale coherent structures in the flow, and (b) an observable that quantifies in a geometric sense and with that the intuitive impression on how Lagrangian particles move in different turbulent flows. }

The statistics of curvature, $\kappa$, and torsion, $\tau$ of Lagrangian particle trajectories have been calculated for homogeneous and isotropic turbulence \cite{Braun,Xu,Scagliarini} and Rayleigh-B\'enard convection \cite{Alards2017}. Across all datasets, the same seemingly universal form for the probability density functions (PDFs) of curvature and torsion have been found, with low-curvature tails proportional to $\kappa$ and high-curvature tails proportional to $\kappa^{-2.5}$, and low-torsion tails proportional to $\tau^0$ and high-torsion tails proportional to $\tau^{-3}$. The power laws for governing both tails of the curvature PDFs and that describing high-torsion events can be derived assuming independent Gaussian statistics of velocity and acceleration \cite{Xu,Alards2017}, which is not the case for turbulent flows.  
That is, curvature and torsion statistics are seemingly insensitive to details of the flow and in particular to at least low or moderate levels of anisotropy.

\moritz{
Based on these results, we anticipate that the presence of turbulent superstructures will not alter curvature and torsion statistics, hence a more refined observable is required to provide information on how large-scale coherent flow structures or a mean flow alter the geometry of Lagrangian particle trajectories.  To do so, we introduce the curvature vector. It distinguishes between different spatial directions and its statistics allow a physical interpretation of how Lagrangian particles move in different turbulent flows, depending on the large-scale structure of the flow.
}

We find that curvature and torsion PDFs for Rayleigh-B\'enard convection (RBC) and von K\'arm\'an
datasets can be mapped to a master curve after appropriate re-scaling with
respect to the Taylor-scale Reynolds number. This may have been anticipated for
the curvature since the bulk curvature statistics in RBC are the same as in
homogeneous isotropic turbulence \cite{Alards2017}.
Here, we extend
these results to torsion statistics. This assessment is corroborated by the
results for the zero-pressure-gradient (ZPG) boundary layer in the logarithmic region, where the torsion
statistics agree with those of the aforementioned datasets. For curvature, the strong unidirectional flow in the ZPG turbulent boundary layer suppresses high curvature events, and low-curvature events become more likely compared to RBC and von K\'arm\'an flow.  In
contrast, the statistics of the curvature vector prove sensitive to even low
levels of anisotropy.  
As such, we demonstrate the curvature vector to be a
useful observable to provide a quantification of anisotropy in the Lagrangian
frame of reference. 

The paper is organised as follows. Section \ref{sec:geometry} outlines the
required background in the differential geometry of space curves in the context
of Lagrangian particle trajectories.  The experiments, collected datasets and
analysis methods are described in section \ref{sec:methods}.  Section \ref{sec:results} contains a summary and comparison of velocity, acceleration, 
curvature and torsion statistics for the different datasets considered here. 
Our main results are presented in sec.~\ref{sec:curv_vector}, which focuses on the statistics of the curvature vector.  
We conclude with a summary of our results and provide 
suggestions for further research in
section \ref{sec:conclusions}.  

\section{Curvature and torsion of space curves} \label{sec:geometry}
\moritz{We consider a Lagrangian particle trajectory $\bm{x}(t)$ in a flow geometrically as given by the motion of a particle along a differentiable curve embedded in three-dimensional space. The Frenet-Serret formulae describe such curves through the respective rates of changes of the tangent vector $\bm{T}$ to the curve, the vector normal to it, $\bm{N}$ and the binormal vector $\bm{B} = \bm{T} \times \bm{N}$ with respect to the arclength $s$ of the curve
\begin{align}
\label{eq:FS-formulae}
\frac{d}{ds}\bm{T} & = \kappa \bm{N} \ , \\
\frac{d}{ds}\bm{N} & = -\kappa \bm{T} + \tau \bm{B} \ , \\
\frac{d}{ds}\bm{B} & = -\tau \bm{N} \ , 
\end{align}
where $\kappa = |d/ds \, \bm{T} |$ is the curvature and $\tau = |d/ds \, \bm{B} |$  the torsion. Since $\bm{T}$ is the tangent vector to the curve with respect to arclength, its rate of change will always be perpendicular to it, hence $\bm{N}$ is normal to $\bm{T}$. }

For Lagrangian trajectories it is useful to express curvature and torsion as time derivatives of the particle position, that is velocity and acceleration, instead of in terms of the arclength of the curve
\begin{align}
    \kappa & =  \frac{|\bm {u} \times \bm{a}|}{| \bm{u}|^3} = \frac{|\bm{a}_n|}{|\bm{u}|^2} \ , \label{eq:curv} \\
    \tau & = \frac{\bm{u} \cdot (\bm{a} \times \dot{ \bm{a}}) }{(\bm{u} \cdot \bm{u} )^3 \kappa^2} \ ,
    \label{eq:tors}
\end{align}
where $\bm{u}$ is the instantaneous velocity, $\bm{a}$ the 
instantaneous acceleration and $\bm{a}_n$ the acceleration component normal to the velocity vector, and $\dot{\bm{a}}$ denotes the total time derivative of the acceleration.

As most flows in nature or engineering applications are not statistically homogeneous and
isotropic, a measurement that reveals these broken statistical
symmetries is needed. Curvature and torsion are global properties that measure
the total instantaneous change in the shape of a trajectory. By construction,
no information about the contributions to the total curvature stemming from the
rate of change of the tangent vector to the curve in each coordinate direction
is retained. In order to obtain such information, in what follows we define the {\em curvature vector}, and propose to 
consider the statistics of its projections onto coordinate directions defined by the experimental configurations.

\subsection{Curvature vector}
As introduced above, the curvature is defined as the magnitude of the rate of change of the tangent vector with respect to the arclength of the curve. In the context of
anisotropy and a fixed coordinate system determined by the experimental
apparatus, one may consider the rates of change of the tangent vector to a
Lagrangian trajectory in the respective coordinate
direction, expressed in terms of time derivatives
\moritz{
\begin{equation}
    \frac{d}{ds} \bm{T} =  \dot{\bm{T}} \frac{dt}{ds} = \frac{1}{|\bm{u}|} \dot{\bm{T}} \ .
\end{equation}
}

\begin{figure}
    \centering
    \includegraphics[width = 0.8\textwidth]{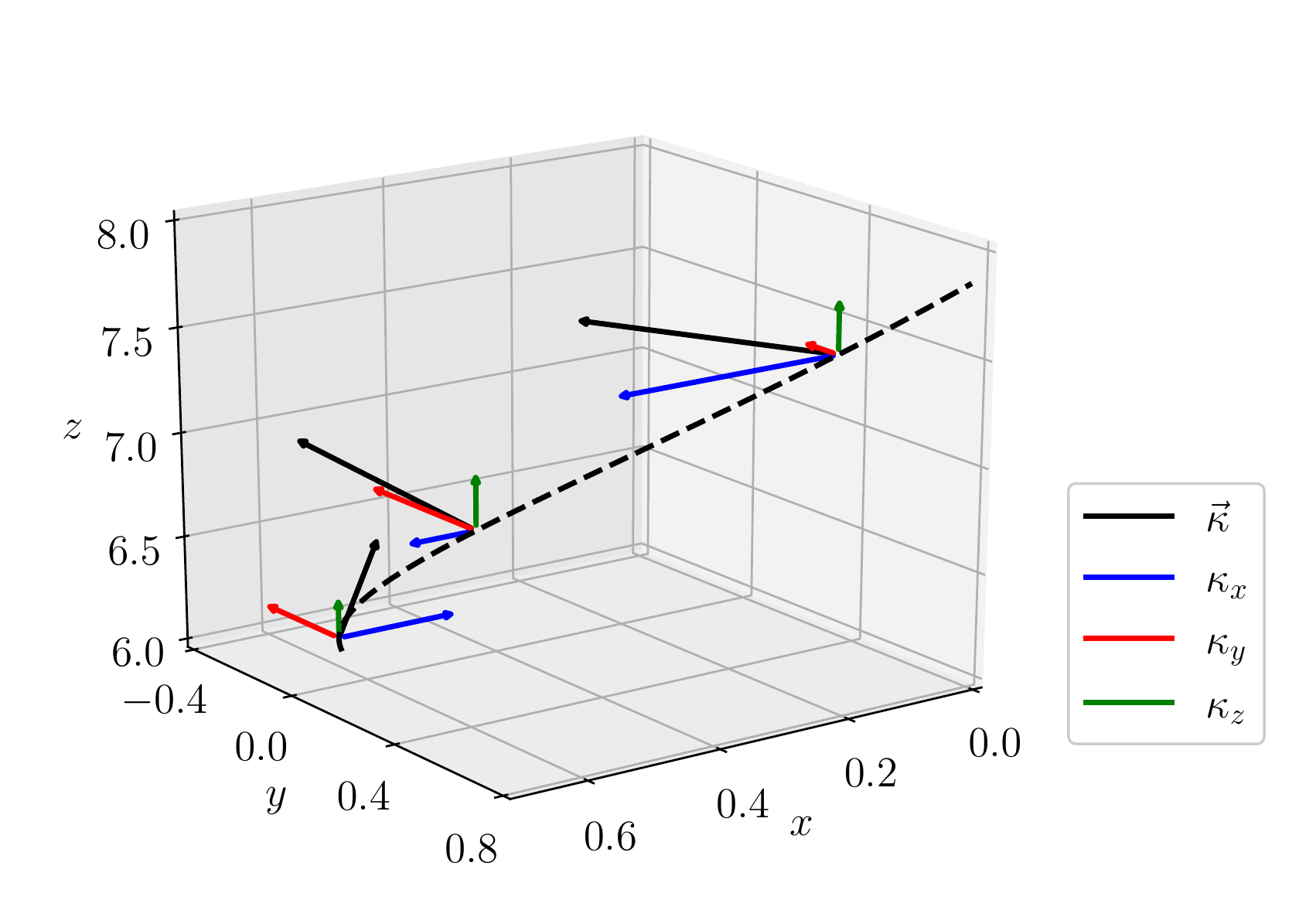}
    \caption{Curvature vector of a trajectory and projections onto the $x$-, $y$- and $z$- directions. The dashed line represents a space curve, the colors indicate the different directions, and curvature vector, as defined in eq.~\eqref{eq:signed-vec}, is shown in black.}
    \label{fig:traj}
\end{figure}

\moritz{
Inspired by the definition of the signed curvature for planar curves and since the arclength derivative of the tangent vector is always perpendicular to the tangent vector, we define the curvature vector as the cross product between $\bm{T}$ and $d/ds \, \bm{T}$ 
\begin{equation}
    \bm{\kappa} = \bm{T} \times \frac{d}{ds} \bm{T} 
    = \bm{T} \times \frac{1}{|\bm{u}|} \dot{\bm{T}} 
    = \frac{\bm {u} \times \bm{a}}{| \bm{u}|^3}\ , 
    \label{eq:signed}
\end{equation}
and its absolute value is given by the expression for the curvature in eq.~\eqref{eq:curv}.
}
The curvature vector, visualised in fig.~\ref{fig:traj} for an example trajectory, is perpendicular to the velocity $\bm{u}$ and the acceleration vector $\bm{a}$, and  
 is closely related to the bi-normal vector as part of the Frenet-Serret formulae. For planar curves, where $\bm{\kappa}$ is always perpendicular to the plane the
curve lies in, it corresponds to the signed curvature. 
To probe the effect of
large-scale coherent flow structures and more generally anisotropy on the
geometry of Lagrangian particle trajectories, we will discuss the statistics of
the projection of this vector onto the $x$-, $y$- and $z$-directions,
respectively,
\begin{align}
    \kappa_x & = \frac{1}{|\bm{u}| ^3} u_y a_z - u_z a_y \ , \nonumber \\ 
    \kappa_y & = \frac{1}{|\bm{u}| ^3} u_z a_x - u_x a_z \ , \nonumber \\ 
    \kappa_z & = \frac{1}{|\bm{u}| ^3} 
    u_x a_y - u_y a_x \ .
    \label{eq:signed-vec}
\end{align}
The curvature vector quantifies what physical intuition would tell us about the geometry of particle trajectories in configurations with a strong mean flow. For instance, by eq.~\eqref{eq:signed-vec} a strong unidirectional flow should lead to
smaller changes in the tangent vector to the curve in the mean flow direction
and hence one may expect a higher probability of small values of the curvature
in this direction compared to the other coordinate directions.  We will see in sec.~\ref{sec:curv_vector}
that this is indeed the case. 

Finally, considering again the expressions for the components of the curvature vector given in eq.~\eqref{eq:signed-vec}, we note that the same argument that determines the shape of the right tail of the curvature PDF \cite{Xu} may be applied here to the PDFs of the curvature vector components. 
 
\moritz{ 
A power-law governing the left tails of the curvature vector components can be derived by noting that each component of the curvature vector can be interpreted as the (signed) curvature of a 2D space curve. We now apply the same arguments proposed in Ref.~\cite{Xu} to determine the asymptotics of the  left tails of the full 3D curvature PDF to the 2D case.  
That is, we assume that the component PDF scale with the normal acceleration in the limit $\kappa_i \to 0$, where $i$ stands for $x, y$ or $z$, and further assume Gaussian statistics for the single acceleration component normal to the -- now 2D -- velocity. The leading order term is then constant.   
In summary, we may expect the left tails of the PDF of the curvature vector components $\kappa_i$ to scale as $\kappa_i^0$. We will see that this is indeed the case.   }

\section{Methods and Data} \label{sec:methods} 

The analysed data consists of a turbulent von Kármán flow,
Rayleigh-Bénard convection at two different Rayleigh numbers and a ZPG turbulent
boundary layer. Key properties and observables of the experiments are provided in
table \ref{tab:flow-properties}, and descriptions of the
different experimental configurations and obtained datasets are provided below. For all datasets, a global
coordinate system is used, that is the standard Rayleigh-Bénard convection
setup where $x$- and $y$-directions define the horizontal plane and $z$ is the vertical direction normal to the heated bottom plate. Therefore the coordinate system for the boundary layer
here has its wall-normal direction in $z$-direction rather than the commonly
used $y$-direction.  \moritz{The coordinate system for the von-Kármán flow is defined to have the propellers along the $z$-axis.}
In all cases, the experimental error is estimated as
noise using a spectral analysis of the particle positions. 
The amplitude spectrum falls of with a specific slope, until it turns towards a 
constant value for high frequencies. \moritz{The frequency at which this transition occurs, is called the crossover frequency $f_c$ and the} positional error $\Delta x$ of the raw particle tracks 
can be estimated as the height of this constant level \cite{errorestimation}.

\subsection{\moritz{Shake-The-Box algorithm}}

\moritz{All present data was obtained by applying Lagrangian Particle Tracking \cite{Schanz23}. In particular, the Shake-The-Box (STB) \cite{stb} algorithm was used to derive long particle trajectories
from the time-resolved projections of a dense field of illuminated flow tracers onto several cameras.
STB overcomes the limitations in seeding density of prior particle tracking methods by combining the use of
an iterative particle triangulation approach and a temporal predictor/corrector scheme.
More precisely, STB employs advanced Iterative Particle Reconstruction (IPR) \cite{Wieneke_2013,Jahn21} to determine 3D particle positions.
From the reconstructed 3D positions of the first few time-steps in a series, short trajectories are extracted by searching for low-acceleration combinations.
These first tracks are then extended (predicted) to the following time-step and an image matching scheme is applied to correct for the occurring prediction errors prior to any reconstruction step.
This 'shaking' 
results in a very reliable extension of known trajectories, while new ones are continually extracted and added by further application of IPR on the residual images until convergence after typically 10-15 time steps is reached. At this state, almost all trajectories inside the measurement volume are known and can be predicted and corrected in all subsequent time steps, while only those few particles newly entering the volume per time step need to be reconstructed by IPR and further added to the tracking system.
The complete process allows the tracking of particles in high numbers (on the order of 100.000 particles per megapixel camera resolution), while avoiding the generation of false (ghost) tracks.}

\begin{table}[]
    \centering
    \begin{tabular}{|c|c|c|c|c|}
        
        & vK & RBC I & RBC II & BL \\
        \hline
        $Re_{\lambda}$ & $270$ & 14\moritz{7}& 18\moritz{6}& \moritz{108}\\ 
        $Re_{\tau}$ & -- & -- & -- & 2295\\
        $\tau_{\eta} [s]$ & $0.013$ & $0.35 $ & $0.18 $ & \moritz{$0.0023$} \\

        $f [Hz]$ & $1250 $ & $20 $ & $30$ & $1000$\\

        $\tau_{\eta} \cdot f $ & 16.25 & 7 & 5.4 & \moritz{2.3}\\

        $\eta [mm]$ & $0.1 $ &$2.3$ & $1.7$ & \moritz{$0.186$}\\

        $\Delta x [\mu m]$ & 3 & \moritz{22} & \moritz{24} & 30 \\

        $\eta/ \Delta x$ & 33 & 96 & 70 & \moritz{6.2}\\ 

        \moritz{$f_c$} &\moritz{$(0.15,0.15,0.18)$ }& \moritz{$(0.18,0.18,0.18)$ }& \moritz{$(0.22,0.22,0.22)$ }& \moritz{$(0.25,0.2,0.25)$} \\

        

        \moritz{$St [10^{-3}]$} & \moritz{0.1} & \moritz{$0.80 \pm 0.17$ }&  \moritz{$0.80 \pm 0.17$} & \moritz{10}\\

        $f_p [Hz]$ & $0.5$ &  -- &  -- &  --\\
         
        $\Bar{\bm{U}} [m/s]$ &  -- &  -- &  -- & $(7,0,0)$\\
         
        $Ra [10^8]$ &  -- & $5.25 \pm 0.06$ & $15.3 \pm 0.3 $ &  --\\
         
        $Pr$ &  -- & 0.7 & 0.7 &  -- \\
         
        $\Delta T [K]$ &  -- & $4.03 \pm 0.06 $ & $11.8 \pm 0.2 $ &  -- \\
         
        $\Gamma$ &  -- & 1 & 1 &  --\\
         
        $\lambda_b [mm]$ &  -- & $11.5$ & $8.5$ &  -- \\
         
        $V [cm^3]$ & \moritz{$0.4 \times 0.15 \times 0.4 $} & $ 55^2 \pi \times 110$& $ 55^2\pi  \times 110$& $280 \times 80 \times 25$ \\
        $V_m [cm^3]$ & \moritz{$0.4 \times 0.15 \times 0.4 $} & $ 52.5^2 \pi \times 104.5$ & $ 52.5^2 \pi \times 104.5$ & $100 \times 80 \times $ \moritz{1}\\
         
        $N$ & 92739514    & 1287242 &  12150782 & 99286611 \\
    \end{tabular}
    \caption{Parameters and key observables of the analysed datasets, von Kármán flow (vK), Rayleigh-Bénard convection (RBC) and the ZPG turbulent boundary layer (BL). The Taylor-scale Reynolds number is $Re_{\lambda} = \sqrt{15} (U_{\text{rms}})^2 \eta^2 / \nu^2 $ where $U_{\text{rms}} = \sqrt{1/3\ (\langle u_x^2 \rangle + \langle u_y^2 \rangle + \langle u_z^2 \rangle)}$ is the root-mean-square of the velocity fluctuations in the different directions, $\eta$ the Kolmogorov length-scale, $\tau_{\eta}$ the Kolmogorov time-scale, and $\nu$ viscosity. The friction Reynolds number is $Re_{\tau} = \frac{u_{\tau} \delta }{\nu}$, with $u_{\tau} = \sqrt{\tau_W / \rho}$ where $\tau_W$ is the wall-shear stress, $\delta$ the boundary-layer thickness and $\rho$ the density; 
     $f$ is the camera (sampling) frequency and $\Delta x$ the experimental uncertainty measured spectrally \cite{errorestimation} \moritz{where $f_c$ is the frequency where the spectra becomes constant.}  
     \moritz{The Stokes number $St$ is the ratio of the particles' response time and a timescale of the flow.}
     Flow-specific properties and parameters are the propeller frequency $f_p$ for von Kármán flow and the mean velocity $\Bar{\bm{U}}$ for the boundary layer. For the RBC, we have the Rayleigh number $Ra =g \alpha \Delta T H^3/(\nu \kappa)$, Prandtl number $Pr = \nu/\kappa$, with $g$ the gravitational acceleration, $\alpha$ the isobaric expansion coefficient, $\kappa$ the thermal diffusivity, $H$ the cell height and $\Delta T$ the temperature difference between top and bottom plate. The aspect ratio is denoted by $\Gamma$ and the thickness of the thermal boundary layer by $\lambda_b=H/(2 Nu)$, with $Nu$ the Nusselt number. The volume $V$ is the volume where the trajectories are recorded, $V_m$ the sub-volume considered for our calculations, and $N$ is the number of total number of trajectories, histograms of  trajectories lengths are shown in fig.~\ref{fig:length}. }
    \label{tab:flow-properties}
\end{table}

\begin{figure}
    \centering
    \includegraphics[width = \textwidth]{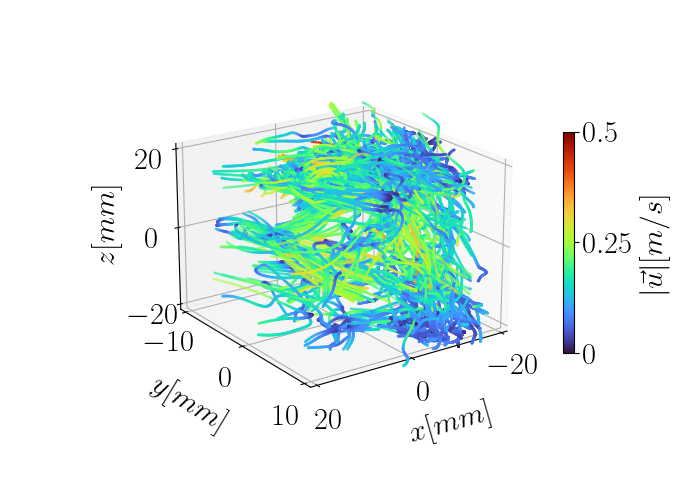}
    \caption{Visualisation of a subset of tracer particle trajectories in von K\'arm\'an flow. The colour bar indicates the absolute value of the velocity. 
    \moritz{updated}}
    \label{fig:visu_vK}
\end{figure}

\begin{figure}
    \centering
    \includegraphics[width = \textwidth]
    {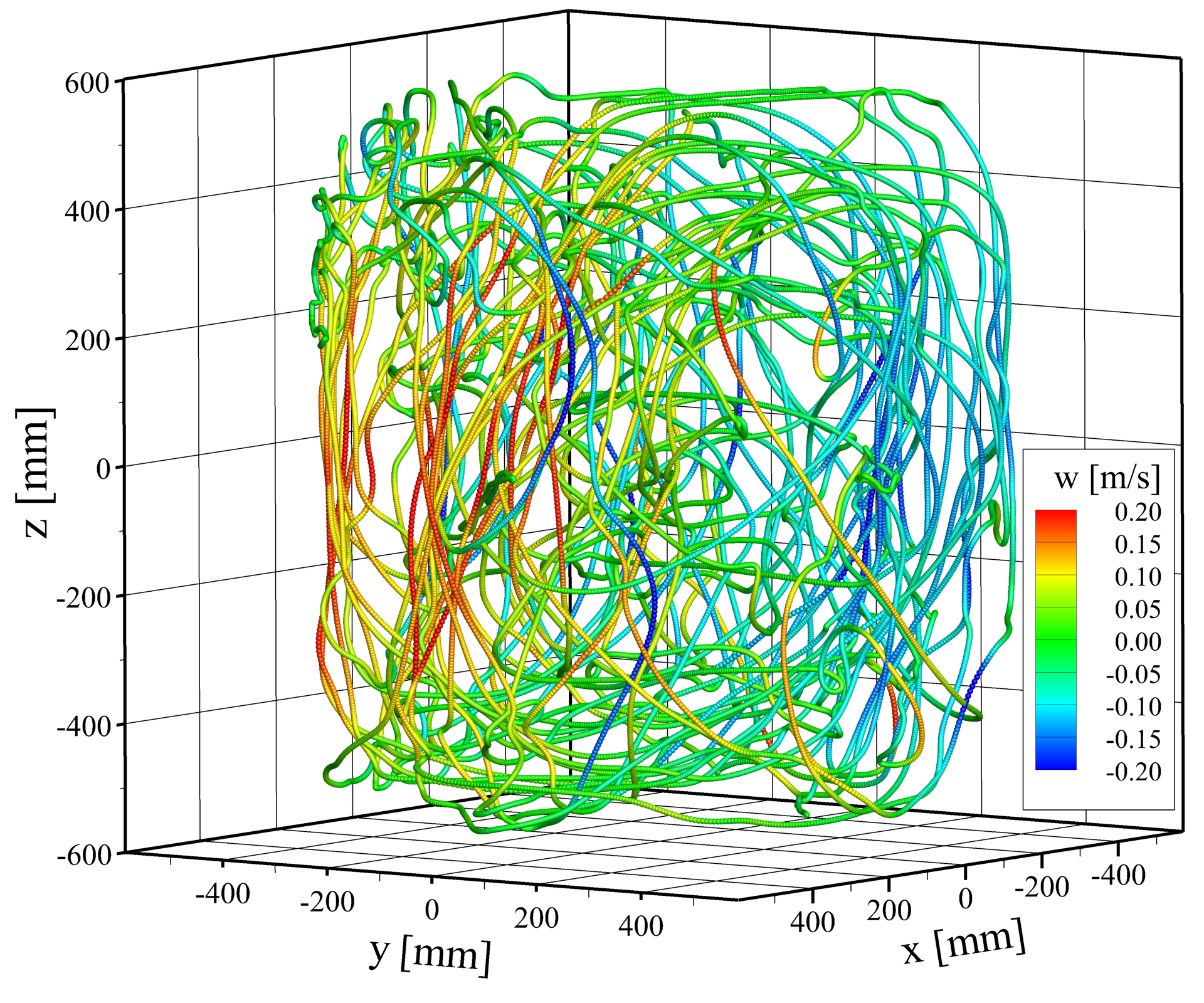}
    \caption{Visualisation of a 
    single long
    tracer particle trajectory in Rayleigh-B\'enard convection at $Ra = 1.53 \cdot 10^9 $. The colour bar indicates the vertical component of the velocity.}
    \label{fig:visu_RBC}
\end{figure}

\begin{figure}
    \centering
    \includegraphics[width = 0.7\textwidth]{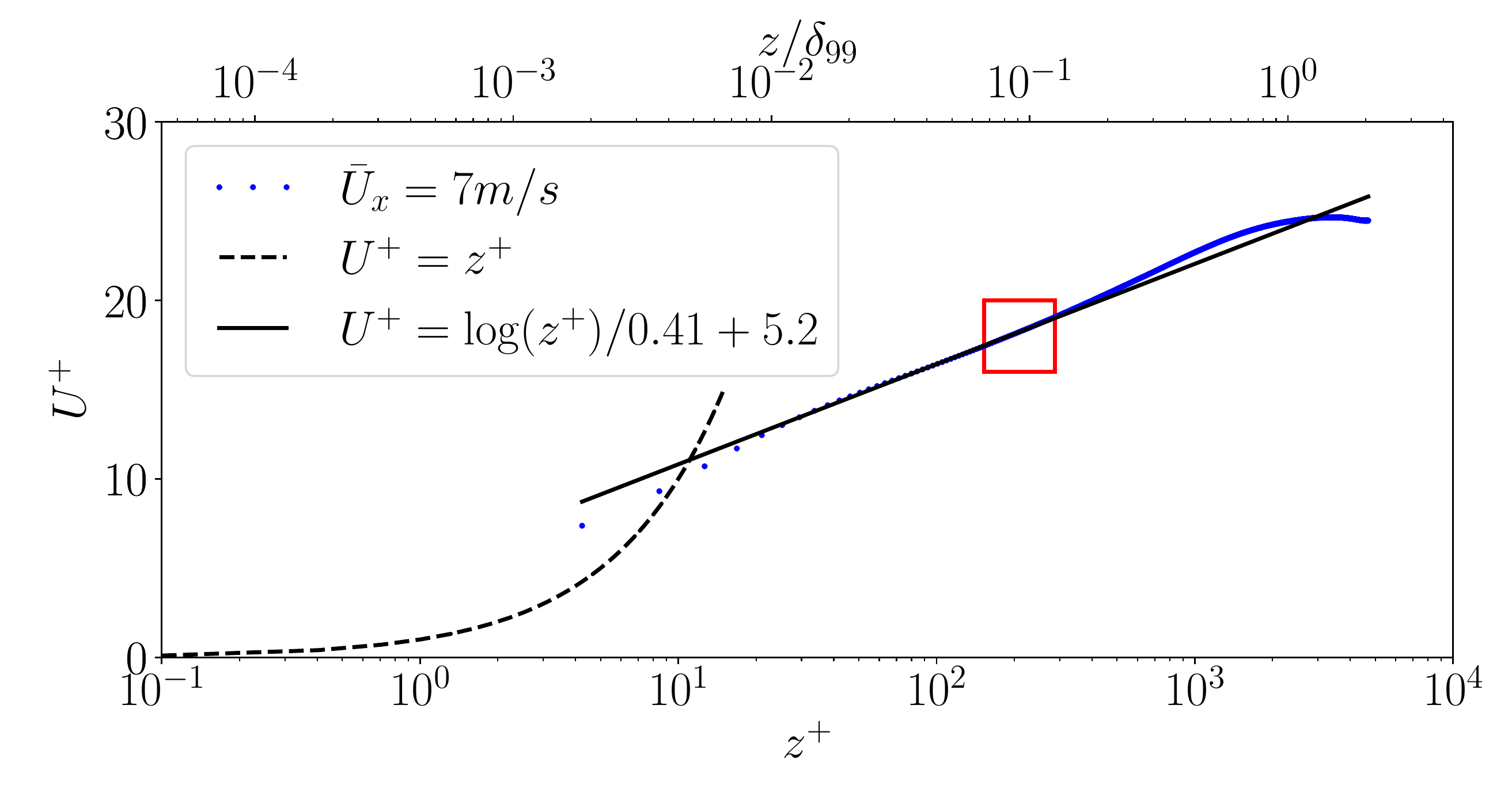}
    \includegraphics[width = \textwidth]{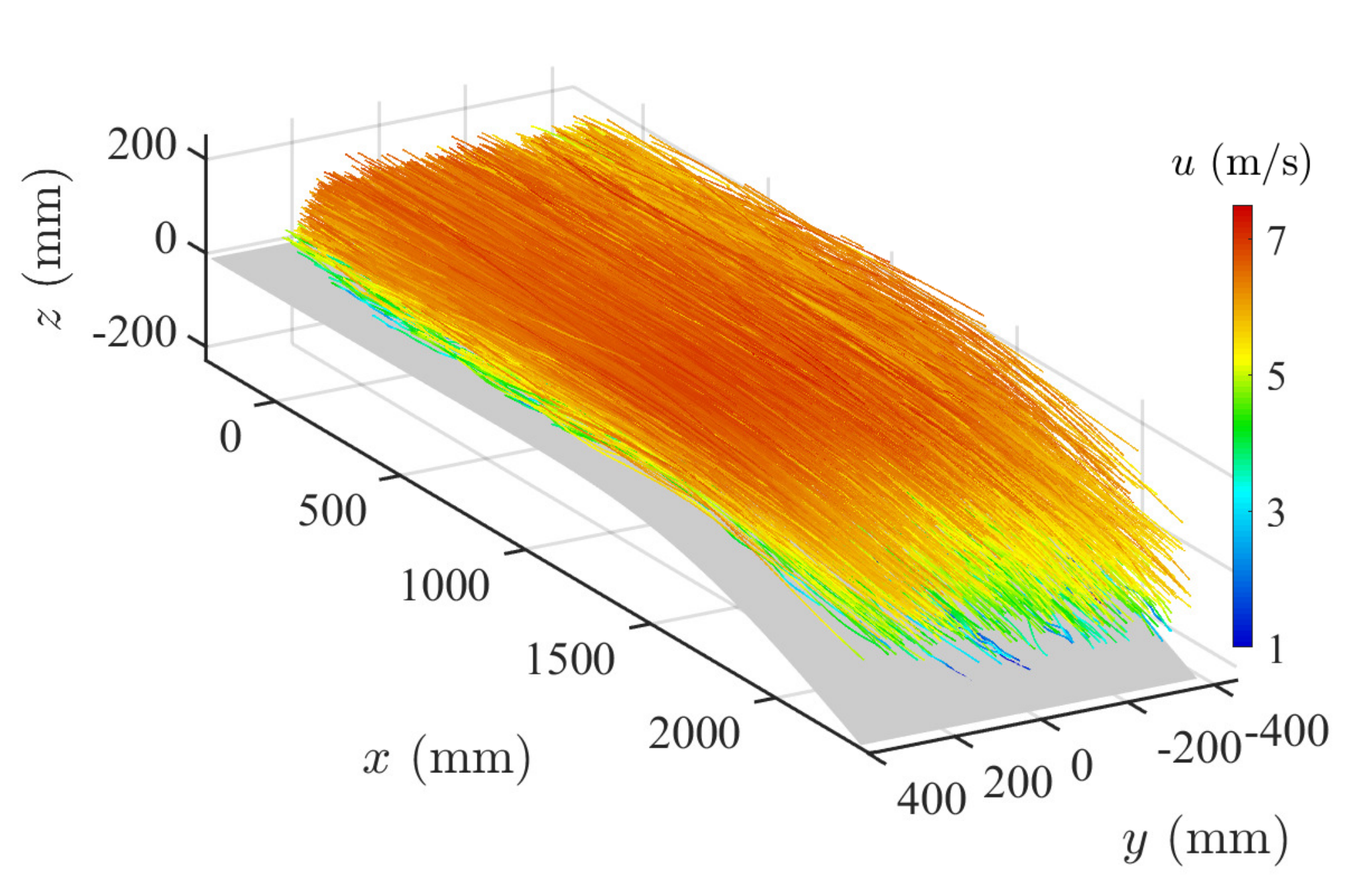}
    \includegraphics[width = 0.7\textwidth]{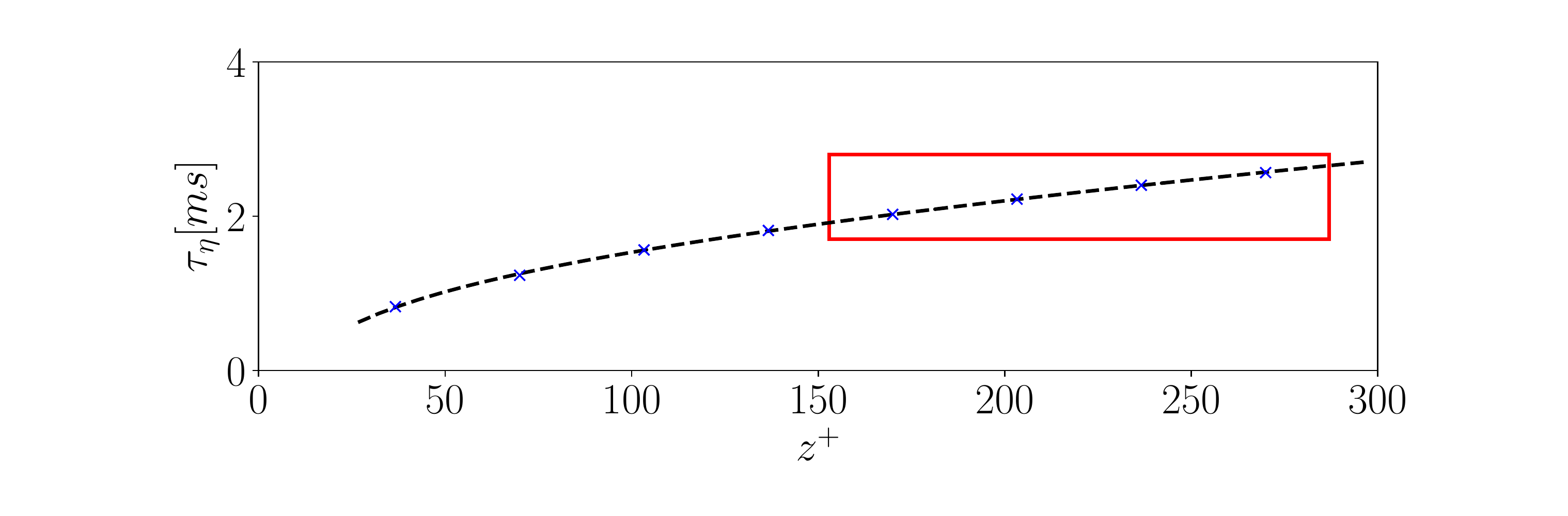}
    \caption{Turbulent boundary layer. Top:
    Stream-wise mean velocity as a function of the distance to the wall, measured in $+$-units where $U^+ = u_x / u_{\tau}$ and $z^+ = z\, u_{\tau}/\nu$ with $u_{\tau}$ the friction velocity and expected velocity profiles. The data clearly follows the logarithmic law from around $z^+=30$ to \moritz{$z^+=400$}. The box indicates the volume used for the analysis presented here, \moritz{that is $z^+ = 153$ to $z^+ =287$}.
    Middle: Visualisation of a subset of tracer particle trajectories within a turbulent boundary layer colored with stream-wise velocity. The ZPG region analysed here extends from $x = 0 {\rm mm}$ until $x = 1000 {\rm mm}$. \moritz{Bottom: Kolmogorov time $\tau_{\eta}$ as a function of the wall normal distance in wall units. The red box indicates the sub-volume used in the analysis.}
    }
    
    \label{fig:visu_BL}
\end{figure}

\subsection{von Kármán flow}

The experiment was carried out by DLR G{\"o}ttingen at the von Kármán facility at the MPI
G{\"o}ttingen \cite{vKdata}. The device consists of two counter-rotating propellers of $500\, mm$ diameter
in a tank filled with water rotating with frequency $f_p = 0.5\, Hz$. That generates approximately homogeneous
and isotropic turbulence in a small volume in the center of the flow chamber. 
The flow is seeded with Dynoseeds TS20 as tracer particles\moritz{, with a Stokes number of $10^{-4}$ effects on the flow can be neglected.} 
\moritz{The particles} are illuminated by a high-repetition speed laser. The camera system recording
the flow consists of four cameras operating at $1250\,Hz$, leading to a high temporal resolution ($\tau_{\eta} \cdot f =16.25$), where $\tau_\eta$ is the Kolmogorov time scale.  
To track the tracer particles the Shake-The-Box algorithm is applied, yielding up to
100000 tracked particles per time step in a volume of approximately \moritz{$40\times15\times40 \,mm^3$}. A visualisation of a subset of particle trajectories \moritz{is} provided in fig.~\ref{fig:visu_vK}.

\subsection{Rayleigh-Bénard convection}

Both datasets of the Rayleigh-Bénard convection were generated at the DLR
G{\"o}ttingen \cite{RBCdata}. The
Rayleigh numbers are $Ra = 5.25 \cdot 10^8$ (RBC I) and $Ra = 1.53 \cdot 10^9$ (RBC II) for the two different datasets.
The experimental set up contains a cylindrical convection cell filled with air at
aspect ratio $\Gamma = 1$ and height $H=1.1\, m$ in $z$-direction and the top and bottom
plate in the $xy$-plane. To ensure constant heating at the bottom, the plate is an electrically heated aluminium plate and for constant cooling at the top, the top plate is water perfused. The flow is seeded with helium filled
soap bubbles as tracers with a diameter of $370\, \mu m$ with an average life expectancy of $326\, s$. \moritz{Their Stokes number is $\approx 8 \cdot 10^{-4}$ which again allows to treat the particles as passive tracers.}
The flow is illuminated by 849 pulsed LEDs placed above the top plate and
synchronised with a system of six scientific cameras operating at $20\,Hz$ and
$30\,Hz$ for the two different cases. 
The tracking of over 500000 tracer particles is again
carried out using the Shake-The-Box algorithm. 
A visualisation of 
single long trajectory is provided in
fig.~\ref{fig:visu_RBC}.
Further details of the experiment and flow visualisations including a video introduction to the experiment are provided in  Refs.~\cite{RBCdata, RBC_data2}. We point out that large-scale motion in form of the large-scale circulation (LSC) is observed in both datasets, see Ref.~\cite{RBC_data2}.

\subsection{Turbulent boundary layer}

The last dataset considered is of a turbulent boundary layer with ZPG \cite{bldata} 
and a bulk flow velocity of $7\, m/s$.
The experiment was conducted in the atmospheric
wind tunnel at the Universität der Bundeswehr München as part of a joint campaign between the University and the DLR. 
The wind tunnel has a $22\, m$ long test section with a $7\, m$ long boundary layer model installed on the side wall several meter downstream from the beginning of the test section. The boundary layer model consists of an S-shaped flow deflection and a downstream straight ramp designed to produce strong adverse pressure gradients up to separation. In between the flow deflections, a $4\, m$ long flat plate is installed over which ZPG conditions are present.

\moritz{The analysis carried out here is restricted to a fraction of the logarithmic layer ($z^+ = 153 - 287$) of the turbulent boundary layer in the ZPG region, as indicated by the box in fig.~\ref{fig:visu_BL}~(top). We only focus on a fraction of the logarithmic layer as it is not possible to define global Kolmogorov scales. We determined the size of the sub volume by balancing the convergence of the data analysis and the assumption of constant Kolmogorov scales (fig.~\ref{fig:visu_BL}~(bottom)). The Kolmogorov length scale varies between $174.24\ \mu m$ and $186.14\ \mu m$ and the time scale between $2.02 \ ms$ and $2.56\ ms$. Details on the method used to calculate the Kolmogorov are given in Ref.~\cite{Falkovich2012}.
} 

The camera system consists of twelve high speed cameras operating at $1000\,
Hz$ recording a volume of approximately $2800 \times 800 \times 250\, mm^3$ in
stream-wise $\times$ span-wise $\times$ wall-normal direction. The ZPG region
was fully recorded with a length of $1800\, mm$ in stream-wise direction. 
Ten high power LED arrays, installed above the wind tunnel, illuminated the flow
containing helium filled soap bubbles (HFSB) as tracers. \moritz{The Stokes number, based on the response time of HFSB ($\approx30\, \mu s$ \cite{Scarano2015}), and the free-stream velocity and displacement thickness, is 0.01. Therefore, we do not expect any inertial effects to play a role and are assuming the particles to be passive tracers.}
Up to 600000 bubbles can be tracked instantaneously within the full volume over time-series of approximately 1382 images using the DLR Shake-The-Box implementation. 
A visualisation of a subset of particle trajectories \moritz{is} provided in fig.~\ref{fig:visu_BL} \moritz{(middle)}.
A more detailed explanation of the experimental set up, calibration and the Lagrangian particle tracking can be found in Ref.~\cite{bldata}. 
We point out that turbulent superstructures in form of coherent structures of considerable stream-wise extent are observed in this dataset, see Ref.~\cite{bldata}, fig.~5.

\subsection{Data processing}

\begin{figure}
    \centering
    \includegraphics[width = \textwidth]{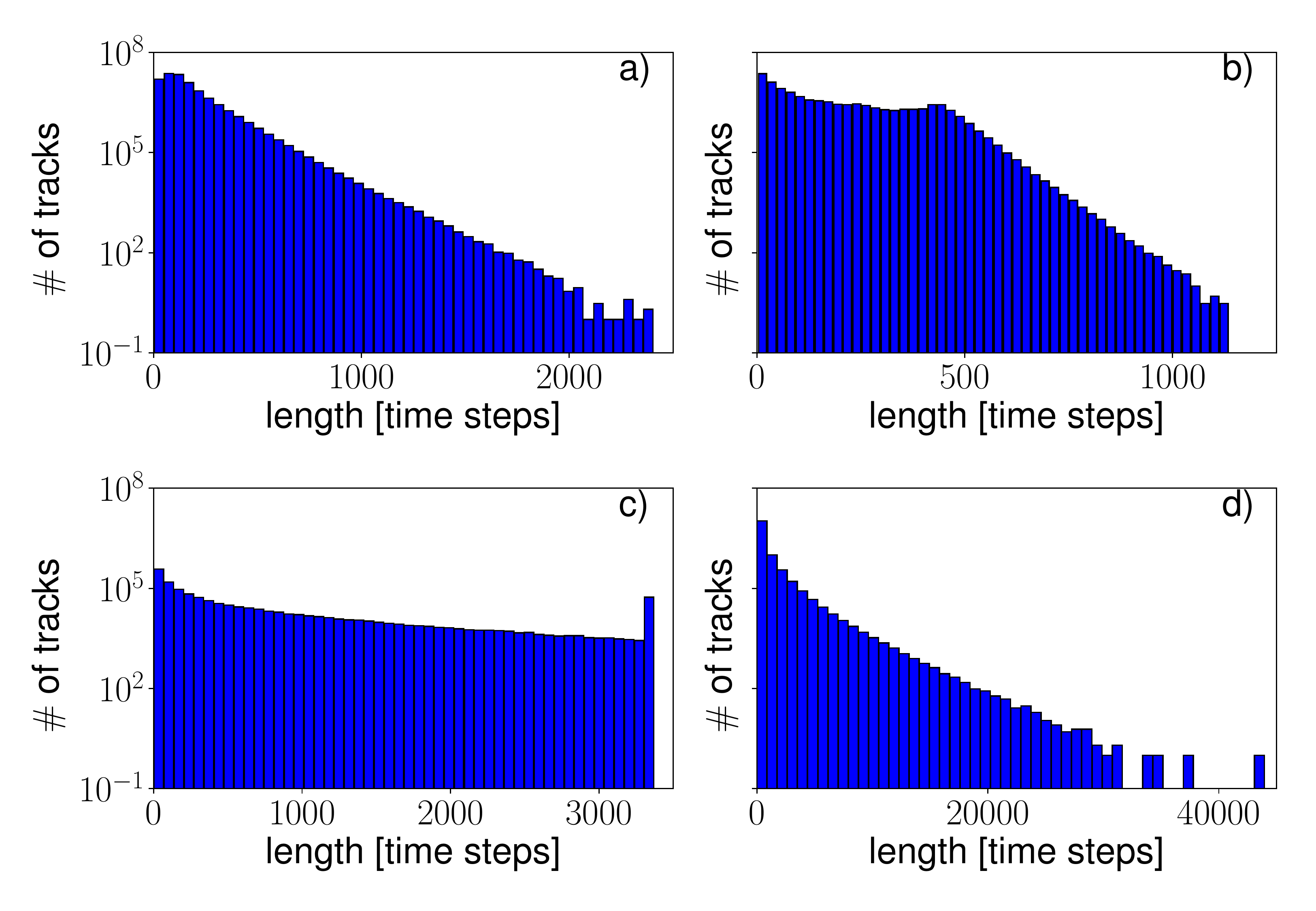}
    \caption{Histograms of trajectory length for a) von Kármán flow, b) boundary layer, c) RBC I and d) RBC II. The histograms are based on the full measurement volume $V$. Trajectories considered for the analysis have a minimal length of 100 time steps. }
    \label{fig:length}
\end{figure}

\begin{figure}
    \centering
    \includegraphics[width=0.9\textwidth]{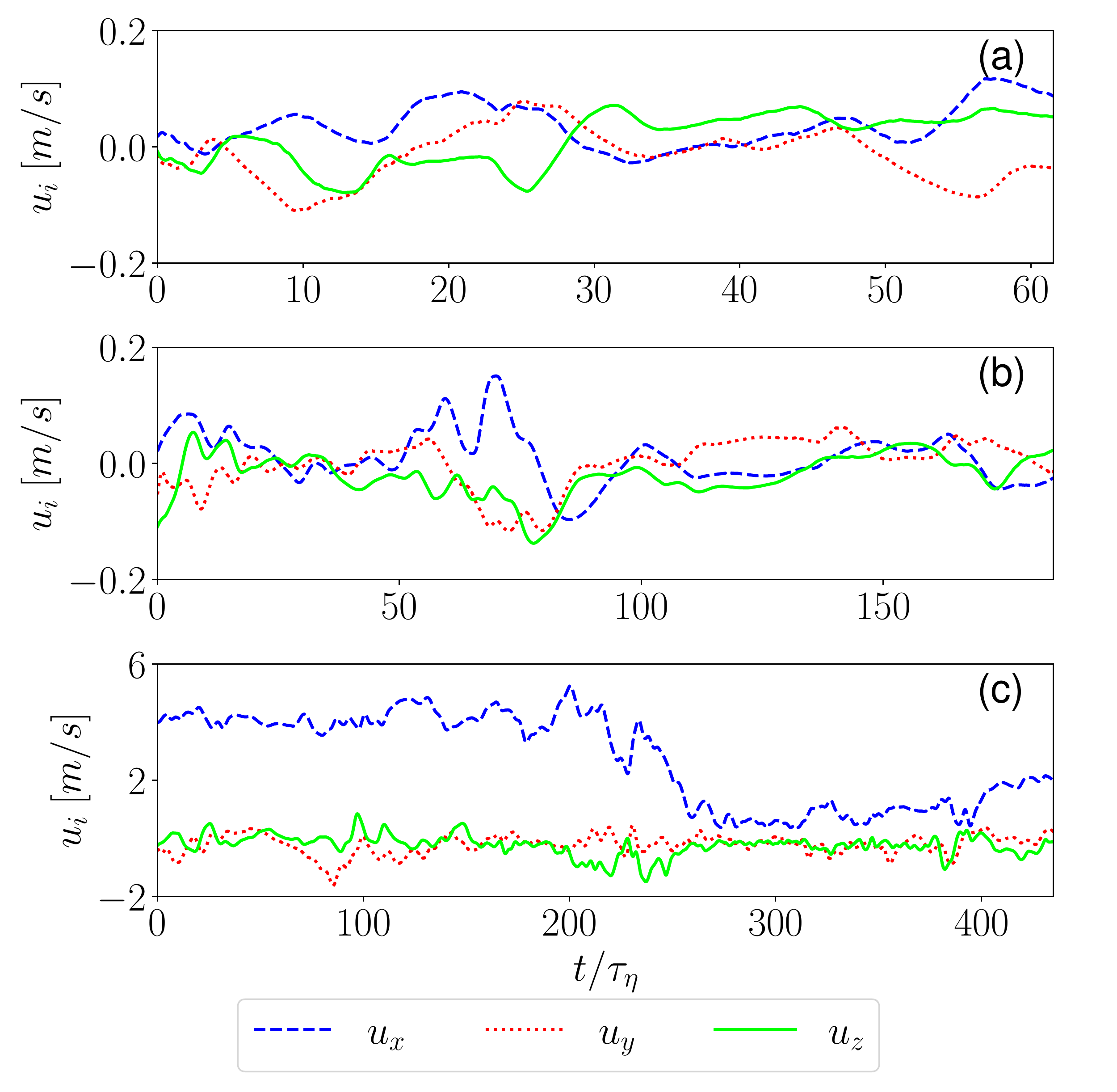}
    \caption{\moritz{Time series of velocity components for (a) von-Kármán flow, (b) RBC II, (c) the turbulent boundary layer.}}
    \label{fig:time-vel}
\end{figure}

\begin{figure}
    \centering
    \includegraphics[width=0.9\textwidth]{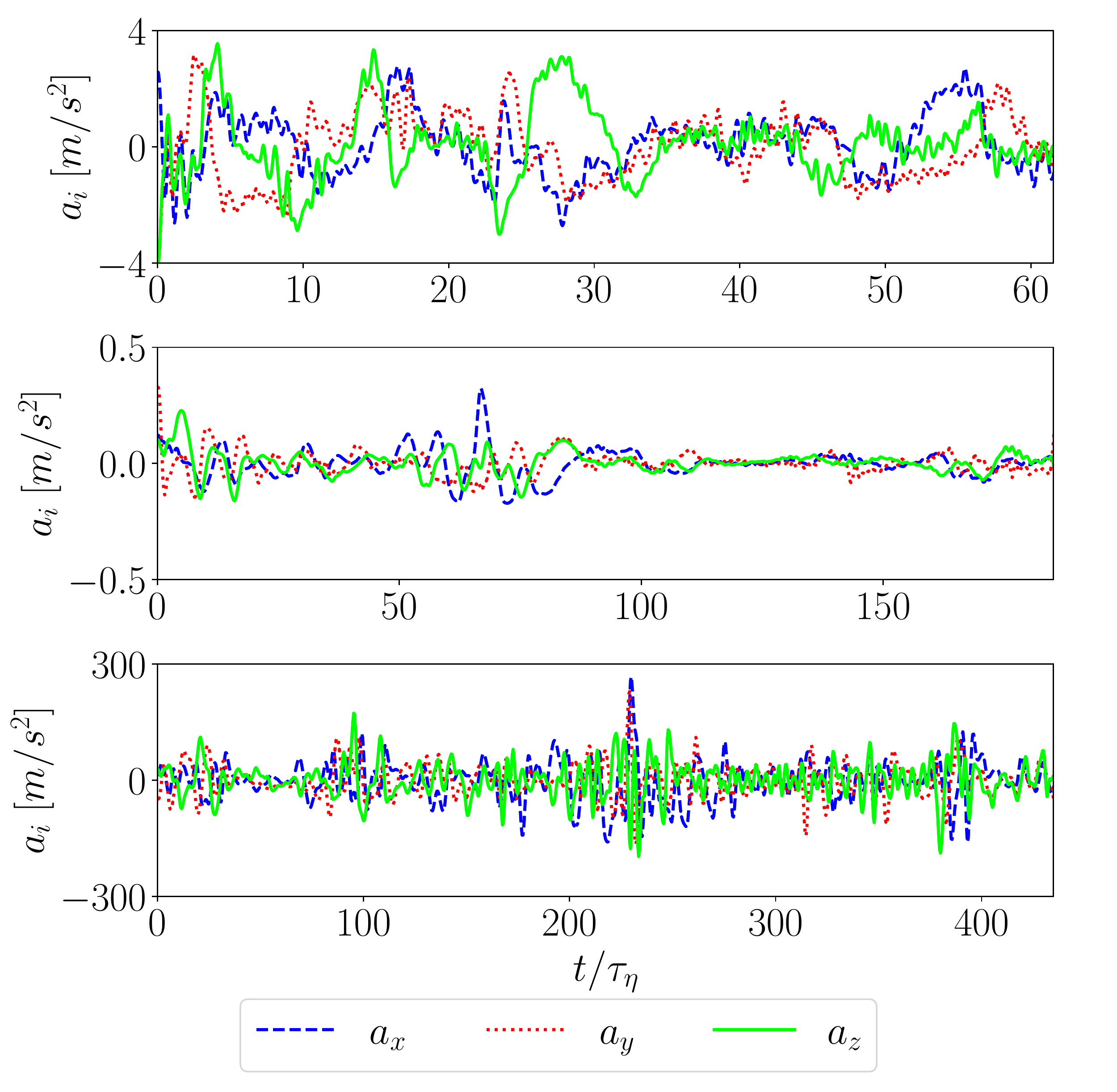}
    \caption{\moritz{Time series of acceleration components for (a) von-Kármán flow, (b) RBC II, (c) the turbulent boundary layer.}}
    \label{fig:time-acc}
\end{figure}

\moritz{To calculate derivatives up to order three, and hence curvature and torsion, accurately, the analysed trajectories should not be too short.}
Histograms of the trajectory
lengths are presented in fig.~\ref{fig:length}. \moritz{As can be seen from the heavy tails of the histograms, we tracked many long trajectories. The mean trajectory lengths are 141 frames for von K\'arm\'an flow, 656 frames for RBC I, 475 frames for RBC II and 155 frames for the turbulent ZPG boundary layer.} We only use trajectories consisting of at least $100$ time steps. 
\moritz{We use the {\em Trackfit} algorithm as introduced by Gesemann et al.~\cite{errorestimation} to fit B-Splines of order three to the particle positions smoothing with an optimal filter length determined by assuming third derivative to be white noise. 
In principle this can become a problem when calculating the torsion. To obtain at least qualitative information on the latter,  
we show time-series of the velocity and acceleration components of a single trajectory of each dataset fig.~\ref{fig:time-vel} and fig.~\ref{fig:time-acc}. Comparing these individually for each dataset, we can see that the velocity fluctuates on time scales larger than for the acceleration, as expected. For all cases we can see that the signal is highly intermittent. However, the time series of the acceleration appears sufficiently smooth to allow a physical interpretation of its derivative, at least for von K\'{a}rm\'{a}n flow and RBC. 
}

\moritz{
To verify that the methods used agree with previous results from the literature, we calculate the acceleration auto-correlation functions: }

\begin{equation}
    R_i(\tau) = \langle a_i(t) \cdot a_i(t+\tau) \rangle \ ,
\end{equation}
where $a_i(t)$ is the instantaneous acceleration in the $i^{th}$-direction. 

The auto-correlation functions are only calculated for trajectories that lie fully
in a selected sub-volume, $V_m$, which for Rayleigh-B\'enard convection excludes thermal and side-wall viscous boundary layers and for the turbulent boundary layer restricts our calculations to the logarithmic region, see table \ref{tab:flow-properties}) for further details. 

For the von Kármán data, a zero-crossing around $2.2 \ \tau_{\eta}$ is expected \cite{Mordant2004,
yeung_pope_1989}, and similarly for Rayleigh-Bénard convection \cite{rbc_correlation}.  Stelzenmuller
\textit{et al.} \cite{bl-correlation} calculated the auto-correlation function in turbulent channel
flow, based on the initial height of the particles. For span-wise and wall
normal direction, the zero-crossing of the auto-correlation function is also found
to be around $2\ \tau_{\eta}$ and shifted to a higher value in stream-wise
direction. We expect the same behaviour for the boundary layer dataset, as the
analysed heights in both cases are where the mean velocity follows the logarithmic law, see also fig.~\ref{fig:visu_BL} (top). \moritz{So far, no physical argument was found, why the auto-correlation functions have the zero crossing around $2\ \tau_{\eta}$ and this is still one of the open questions in turbulence theory.} 

\begin{figure}
    \centering
    \includegraphics[width = \textwidth]{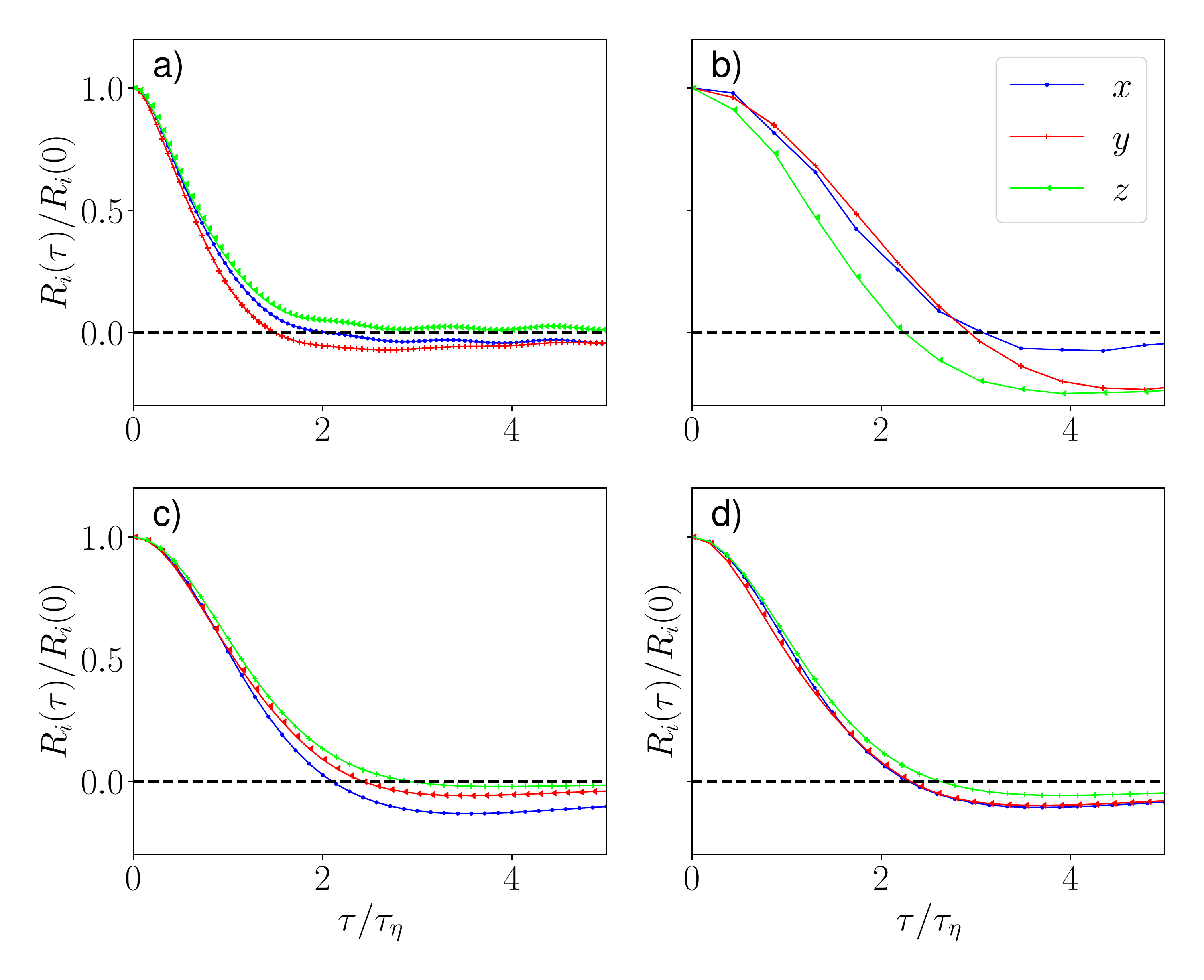}
    \caption{Normalised auto-correlation functions a) for von Kármán-flow with zero-crossing around $2\ \tau_{\eta}$ as expected from analysis in \cite{Mordant2004,yeung_pope_1989}, b) for BL, \moritz{the zero-crossing is between $2.2\ \tau_{\eta}$ and $3.1 \ \tau_{\eta}$}. In c) and d) auto-correlation of RBC with $Ra = 5.25\cdot 10^8$ and $Ra = 1.53 \cdot 10^9$ respectively, both cases show a zero-crossing around $2.2\ \tau_{\eta}$ as found in \cite{rbc_correlation}. \moritz{updated}}
    \label{fig:correlation}
\end{figure}

\moritz{In Figure~\ref{fig:correlation} the correlation functions of the acceleration components are shown. For the von Kármán flow and the two RBC cases the acceleration auto-correlation functions have the expected zero-crossings around $2 \ \tau_\eta$ with slight differences in the different directions. For the boundary layer, the zero-crossing is around $2.9 \ \tau_{\eta} $ and $2.2 \ \tau_{\eta}$ in $y-$ and $z-$directions, respectively. In stream-wise direction, the correlation function crossed zero around $3.1 \ \tau_{\eta}$. This is slightly higher than reported for the pipe flow in \cite{bl-correlation}. What should be noted here is that the temporal resolution is close to the Kolmogorov timescale, and therefore events on this timescales could be filtered out.} 

\section{Velocity, acceleration, curvature and torsion} \label{sec:results}

The following sections present probability density functions (PDFs) of velocity,
acceleration, curvature and torsion  
for the previously described datasets. 
We compare the results with literature as well as between the different datasets, focusing in the first instance on similarities
and differences between the respective curves and model
predictions assuming statistical homogeneity and isotropy \cite{Xu}, and
subsequently also on the effect of the Taylor-scale Reynolds number $Re_{\lambda}$.

\subsection{von Kármán flow}

\begin{figure}
    \centering
    \includegraphics[width = \textwidth]{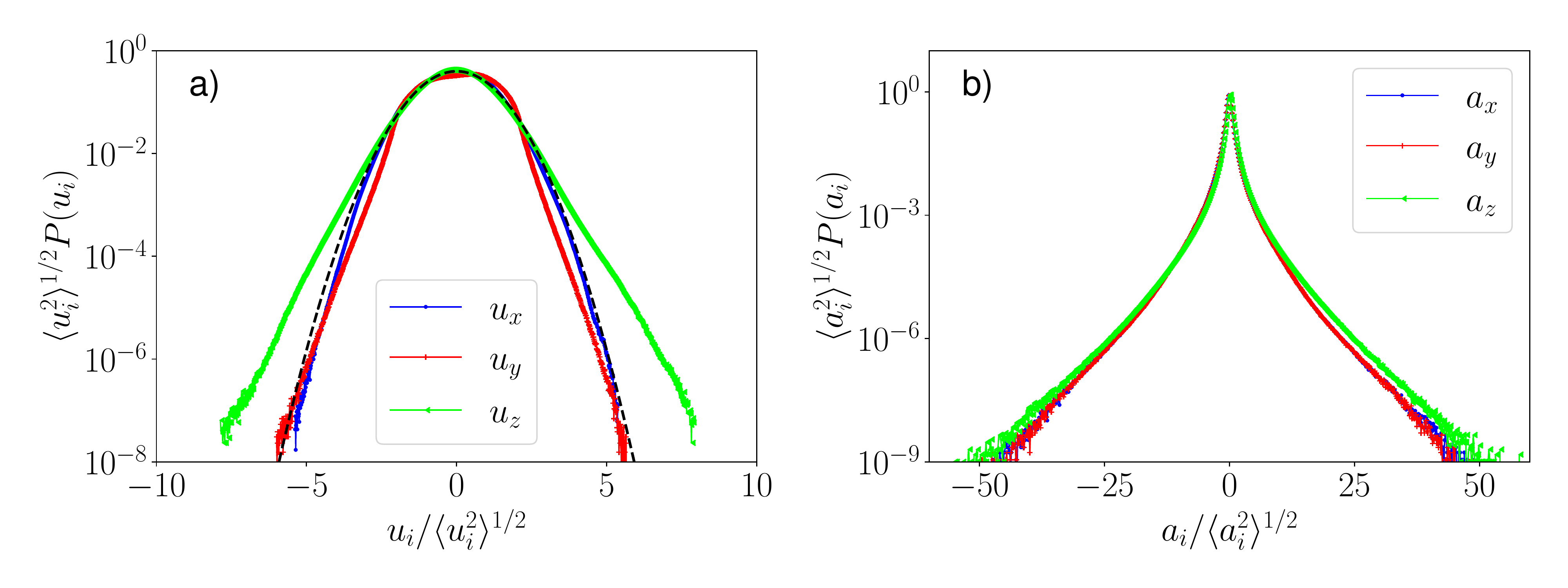}
	\caption{Standardised PDFs of a) velocity, b) and acceleration components for von K\'arm\'an flow on semi-logarithmic scales. 
	The dashed line in a) corresponds to a Gaussian
	with zero mean and a standard deviation of unity. \moritz{updated figure}}
    \label{fig:vk_va}
\end{figure}

Figure \ref{fig:vk_va} a) presents the standardised PDFs of each velocity component of the von Kármán flow at $Re_{\lambda} = 270$
with PDFs of $u_x$ and $u_z$,  $P(u_x)$ and $P(u_y)$, being approximately Gaussian as expected 
and first reported in Refs.~\cite{townsend_1947, Batchelor}. \moritz{In fact, the flatness values of $P(u_x)$ and $P(u_y)$ are 2.77 and 2.33, respectively, indicating slightly sub-Gaussian statistics.} The PDF of \moritz{the velocity component normal to the propellers, $u_z$}, however, 
has super-Gaussian tails \moritz{leading to a flatness value of 3.44}. The latter indicates that extreme velocity fluctuations are more likely in the $z$-direction 
compared with the $x$- and $y$-directions. 
\moritz{This shows that} large-scale fluctuations that break statistical isotropy occur, \moritz{which} is known indeed that 
von Kármán flow is not fully isotropic \cite{vk-anisotropy}. Its large-scale dynamics is dominated by a shearing and a pumping mode, the former being the result of the counter-rotating propellers, while the latter drives fluid first inwards and subsequently upwards towards the propellers through centrifugal pumping, see the visualisations in fig.~11 of ref.~ \cite{Voth2002}. 
\moritz{Furthermore, as the ratio of 
transverse and axial components of the rms velocity of around 1.5 varies only weakly with propeller speed, 
the large-scale dynamics appear not to depend strongly on Reynolds number \cite{Voth2002}.}
By inspection 
and comparison between the standardised PDFs of each acceleration component 
shown in fig.~\ref{fig:vk_va} b), we note that all PDFs have similarly wide tails, 
in qualitative agreement with known results on Lagrangian acceleration statistics in
homogeneous and isotropic turbulence obtained from numerical data \cite{Bentkamp2019,
Biferale, wilczek2, Mordant2004, Mordant_2004_njp} and experimental data
\cite{LaPorta2001, lalescu_wilczek_2018,wilczek2,Mordant_2004_njp}.  For
a quantitative comparison, we consider numerical data at a slightly higher
value of $Re_\lambda = 350$
\cite{Bentkamp2019}, 
where extreme events up to 40 times the standard deviation
occur with probabilities around $10^{-7}$, see figure 3(a) of
Ref.~\cite{Bentkamp2019}. As can be seen from fig.~\ref{fig:vk_va} b), our
results are commensurate with these values.

\begin{figure}
    \centering
    \includegraphics[width = \textwidth]{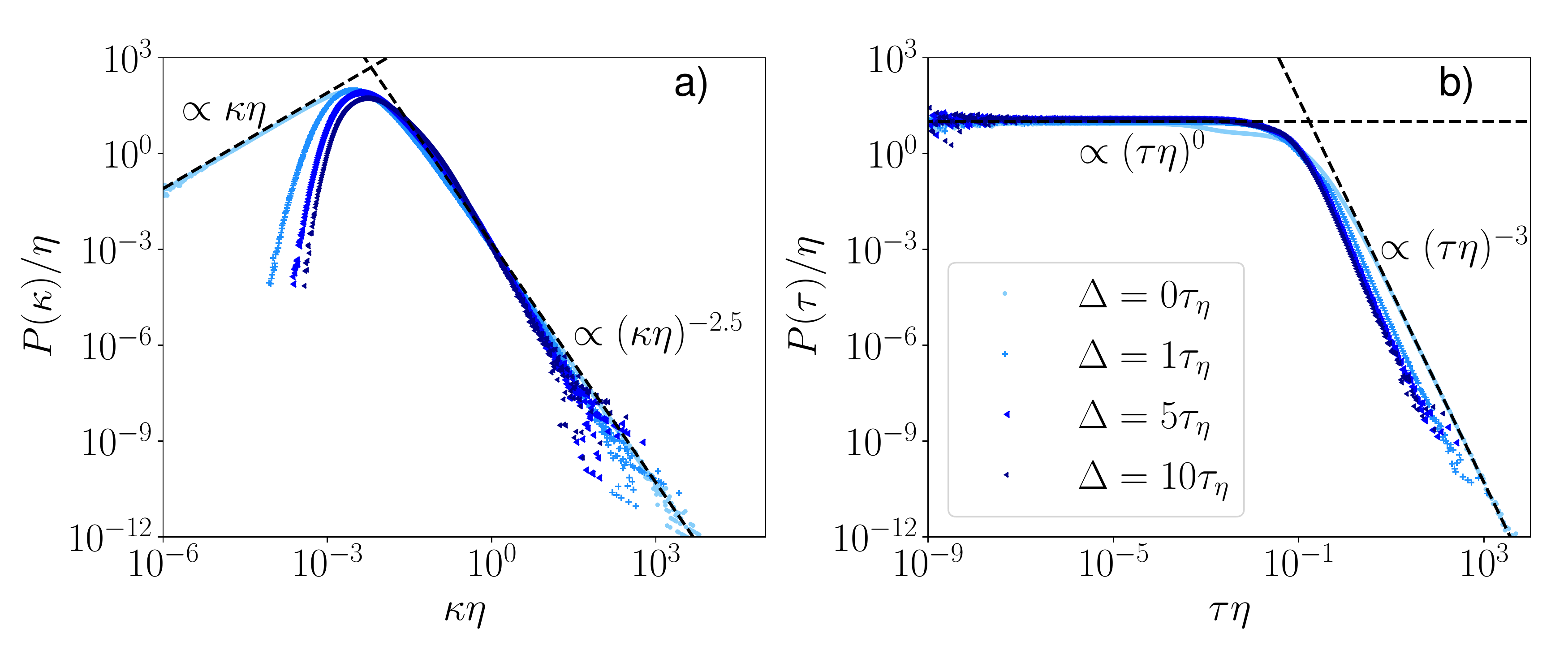}
    \caption{a) Curvature PDFs and b) torsion PDFs for the von Kármán dataset. PDFs of averaged curvature and torsion over different time windows $\Delta$, light blue no averaging, dark blue $\Delta = 10\ \tau_{\eta}$. \moritz{updated figure} }
    \label{fig:vkcurvtors}
\end{figure}

Curvature and torsion PDFs are shown in figs.~\ref{fig:vkcurvtors} a) and b), respectively, 
and we observe the expected power laws for left and right tails of the PDFs 
as in previous work on curvature and torsion statistics in homogeneous and isotropic turbulence using 
either DNS data for curvature and torsion \cite{Braun, Scagliarini} or
experimental data for curvature only \cite{Xu}. That is, the left tail of the curvature PDF that corresponds to 
small-curvature events is proportional to $\kappa \eta$ while the right tail that corresponds to 
high-curvature events scales as $(\kappa \eta)^{-2.5}$. For the torsion PDF we find $P(\tau) \sim (\tau \eta)^{0}$
for low-torsion events and $P(\tau) \sim (\tau \eta)^{-3}$ for high-torsion events. 
These curvature PDF exponents can be derived assuming independent Gaussian statistics for velocity and acceleration 
\cite{Xu}, similar arguments apply to the exponent describing the right tail of the torsion PDF 
\cite{Scagliarini, Alards2017}. Acceleration statistics in developed turbulence are not Gaussian and 
velocity and acceleration statistics are not independent either \cite{Mordant2004}; however, 
using a recently developed decomposition technique of Lagrangian statistics into Gaussian
sub-ensembles \cite{Bentkamp2019}, the PDF exponents can be derived without these assumptions \cite{Hengster_in_prep}.

Rather than connecting curvature with vortical flow structures, Xu \textit{et al.} \cite{Xu} showed that large-scale flow reversals affect curvature statistics and suggest to filter out these events in order to probe the 
more intuitive idea that connects curvature with vorticity. As large-scale flow reversals 
generally occur on short time scales, the filtering was implemented 
by averaging the curvature along a trajectory over a small interval in time. 
Such filtering is also useful to detect correlations between high-normal-acceleration events and vortex filaments \cite{Biferale2005}.

To compare with previous measurements of curvature statistics and extend to torsion statistics, we filtered curvature and torsion along each trajectory for time intervals
of one, five and ten Kolmogorov times with results shown in figs.~\ref{fig:vkcurvtors} a)
and b) alongside the unfiltered case for curvature and torsion, respectively. 
 \moritz{The filtering intervals were chosen to match those in Ref.~\cite{Xu}. Ideally a comparison to the time scales of large-scale flow reversals would be in order, however, we cannot extract this information from our data.}
For the curvature PDFs of the filtered data, the right tail corresponding to large values of the curvature seems to
remain relatively stable, while we observe significant differences between the left tails of the PDFs
associated with small values of the curvature compared with the unfiltered
case. Small values of curvature are becoming increasingly less likely with 
increasing filter scale. Similar results have been reported by Xu \textit{et al.} \cite{Xu}, 
albeit with more profound effects in the high-curvature tail.
For torsion PDFs, we observe the opposite trend, with the low-torsion tail remaining largely unaffected by the filtering while high torsion events become less likely with increasing filter width. Disregarding potential correlations, such behaviour can be explained by inspection of eq.~\eqref{eq:tors} that describes the torsion, where the curvature appears in the denominator. Hence a suppression of low curvature events can plausibly result in a suppression of high torsion events. 

\subsection{Rayleigh-B\'enard Convection}
In what follows we consider velocity, acceleration, curvature and torsion statistics for the two RBC datasets described in section \ref{sec:methods} with Rayleigh numbers $Ra=5.25 \cdot 10^8$ and $Ra=1.53 \cdot 10^9$, respectively. We focus on the bulk by conditioning the statistics on wall-distance in both vertical and horizontal directions resulting in the measurement domains $V_m$ reported in table \ref{tab:flow-properties}. 
 \moritz{For the RBC II case, we estimated that after $2 \ \lambda_b$ the average velocity is 92\% of the maximal velocity. To disregard any effects of the walls, we decided to disregard data at distances of $25.5\ mm = 3\ \lambda_b$. For the RBC I case, we used the same distances, resulting in $2.2 \ \lambda_b$.   }
 
\begin{figure}
    \centering
    \includegraphics[width = \textwidth]{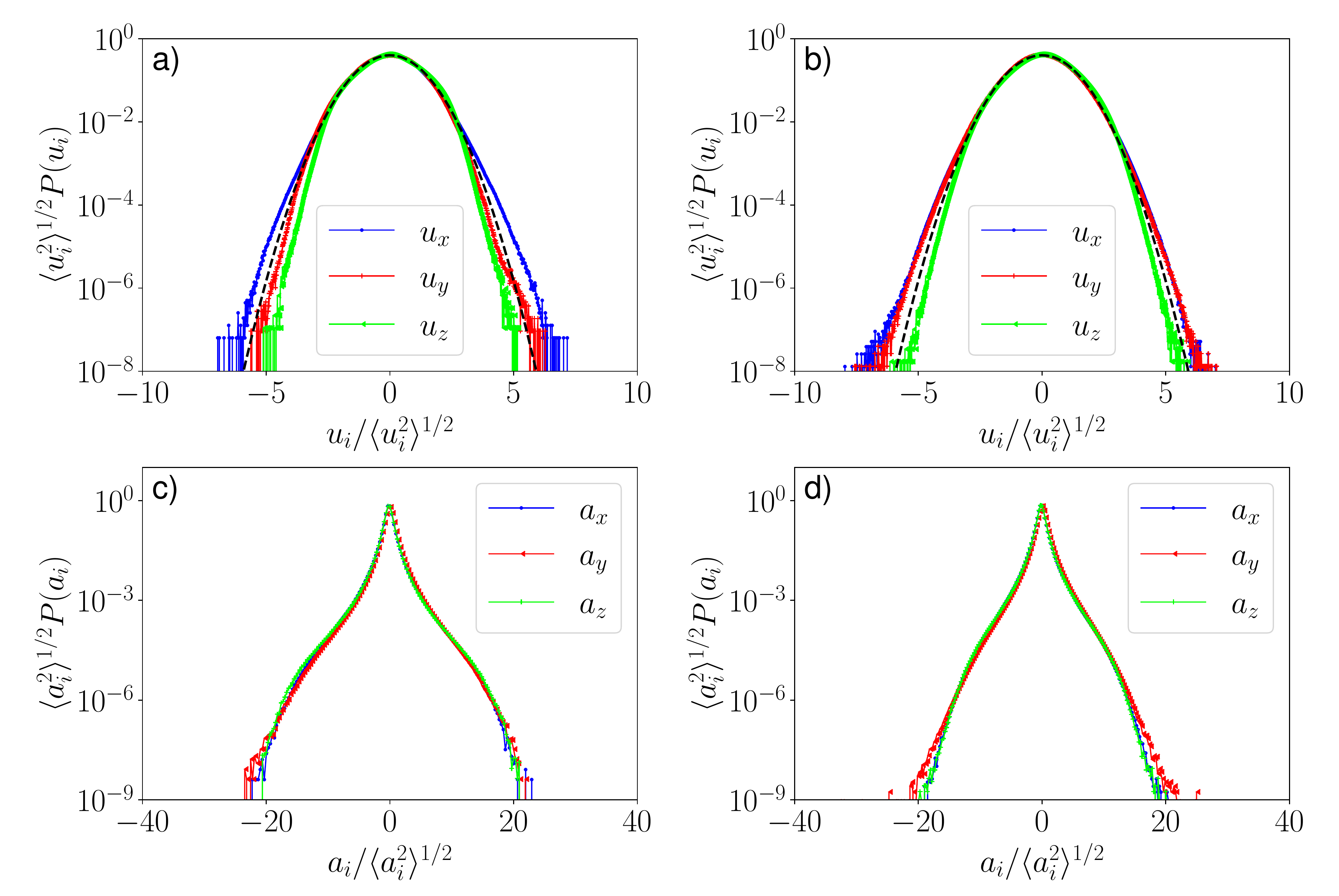}
    \caption{Standardised velocity and acceleration PDFs for Rayleigh-B\'enard convection. Top row: velocity components for a) $Ra = 5.3 \cdot 10^8$ and b) $Ra = 1.53 \cdot 10^9$. The black dashed line corresponds to a Gaussian with zero mean and unit variance. Bottom row: acceleration components for c) $Ra = 5.3 \cdot 10^8$ and d) $Ra = 1.53 \cdot 10^9$.  \moritz{updated}}
    \label{fig:rbc_au}
\end{figure}

Velocity and acceleration statistics for both datasets are presented in fig.~\ref{fig:rbc_au}.  Figures \ref{fig:rbc_au} a), b) present velocity-component PDFs for RBC I and RBC II, respectively. From subfigure a) it can be seen that for the lower-$Ra$ case the PDF of $u_x$ is \moritz{slightly super-Gaussian with a flatness value of 3.18} while those of the remaining directions have approximately Gaussian and slightly sub-Gaussian tails \moritz{with flatness values of 2.92 and 2.84, in $y$ and $z$ direction, respectively}.
For the higher-$Ra$ case, the velocity PDFs in subfigure b) 
\moritz{in the horizontal directions are slightly super-Gaussian and the PDF of the vertical component is} approximately Gaussian.
\moritz{Flatness values for velocity PDFs in $x$-, $y$- and $z$-direction are $3.22, 3.14$ and $3.02$, respectively.} 
The more pronounced sub-Gaussianity in the lower-$Ra$ case compared to the higher-$Ra$ case may be connected to a less perturbed large-scale circulation in the former case. 
Interestingly, the deviations from Gaussianity in both cases are smaller than for the von K\'arm\'an flow. 

Acceleration PDFs for RBC I and RBC II are shown in \ref{fig:rbc_au} c), d), respectively. 
As can be seen by comparison of the data in the two subfigures, the PDFs in \moritz{all} directions are very similar in shape with fluctuations up to 25-30 times the root-mean-square value. 
As $Re_{\lambda} = 270$ is much higher for the von K\'arm\'an data, 
than for the two RBC datasets which corresponds to \moritz{$Re_{\lambda} \approx 147$ and $Re_{\lambda} \approx 186$},  one could have expected wider tails for the von K\'arm\'an acceleration PDFs compared to the RBC datasets. 
The measurements presented here for $Ra = 5.3 \times 10^8$ are commensurate with those reported by Schumacher \cite{Schumacher2009} in the bulk for numerical simulations at $Ra = 1.2 \times 10^8$ in a domain with free-slip boundary conditions on the top and bottom plate and periodic boundary conditions in the transverse directions. 

\begin{figure}
    \centering
    \includegraphics[width = \textwidth]{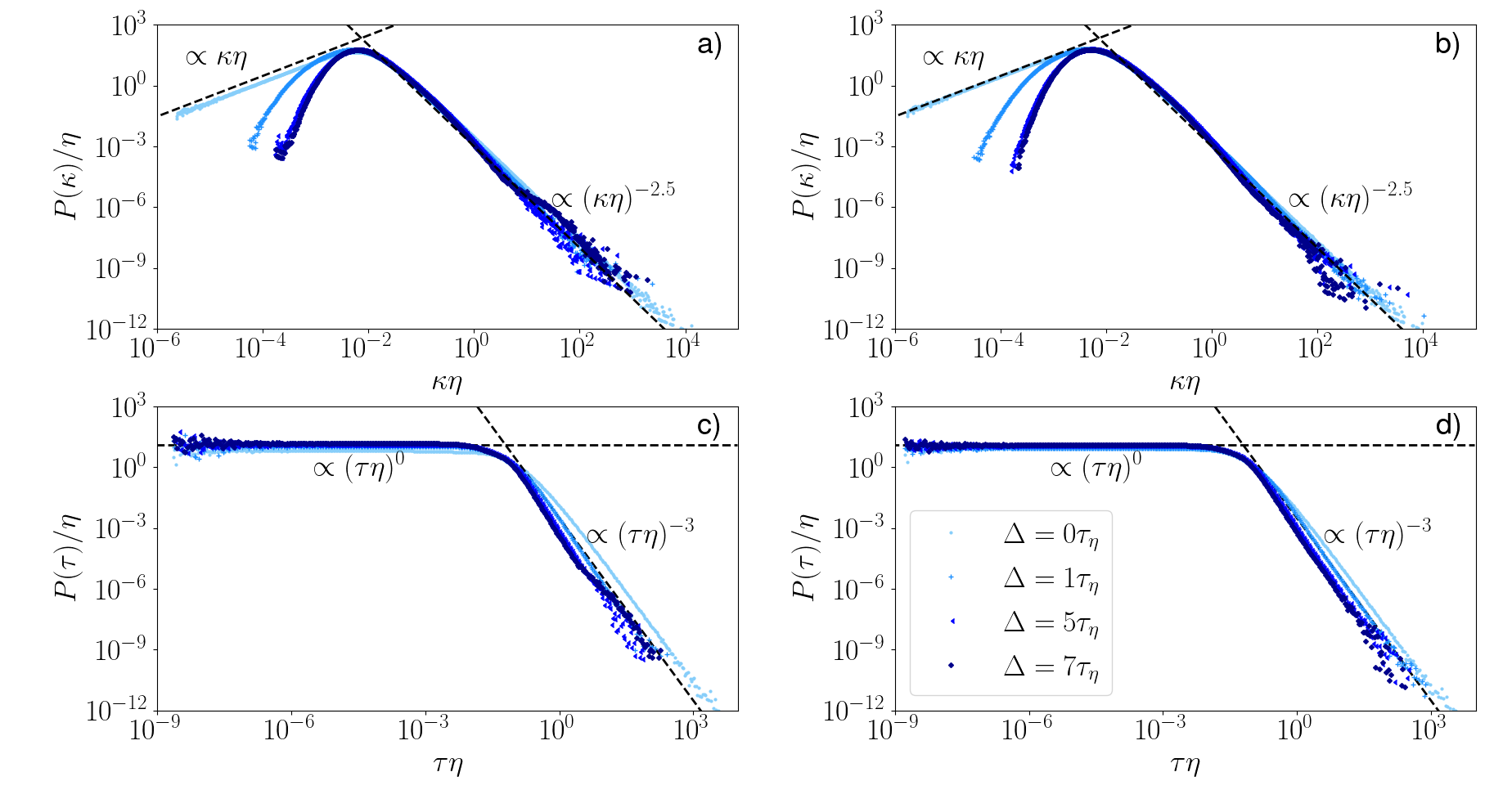}
    \caption{Filtered and unfiltered curvature and torsion statistics for Rayleigh-B\'enard convection, where $\Delta$ is the filter width. Top row: curvature PDFs for a) $Ra = 5.3 \cdot 10^8$ and b) $Ra = 1.53 \cdot 10^9$. Bottom row: torsion PDFs for c) $Ra = 5.3 \cdot 10^8$ and d) $Ra = 1.53 \cdot 10^9$. \moritz{The filter widths are adjusted to ensure filtering on timescales of less than the free-fall time ($\approx 10 \ \tau_{\eta}$).}
     }
    \label{fig:RBCcurv_tors}
\end{figure}

Figure \ref{fig:RBCcurv_tors} presents filtered and unfiltered curvature and torsion PDFs for both datasets. For both observables and both Rayleigh numbers, we find the same power laws as in the case for homogeneous and isotropic turbulence, turbulent von Kármán flow and previous works on non-rotating and rotating Rayleigh-Bénard convection and rotating
electromagnetically forced turbulence \cite{Alards2017}. For the
unfiltered case, the PDF of the non-dimensionalised curvature is linear for
small values and proportional to $(\kappa \eta )^{-5/2}$ \moritz{for high values}, for the torsion the
PDF is constant for small values and the right tail follows a power law with
exponent of -3. Filtering out flow reversals by averaging curvature and torsion along each trajectory leads to a similar behaviour than for the von Kármán flow where the tail for high values of the curvature does not change while for small values of the curvature, the power law changes and {\em vice versa} for the torsion.

\subsection{Turbulent boundary layer}
For the ZPG boundary layer we focus only \moritz{on a sub-volume in the logarithmic region, that is, distances of 
$z^+ = 153 - 287$} from the bottom wall. This dataset differs substantially from the von K\'arm\'an and RBC datasets \moritz{by having }
a strong mean velocity in $x$-direction, see table \ref{tab:flow-properties}.

\begin{figure}
    \centering
    \includegraphics[width = \textwidth]{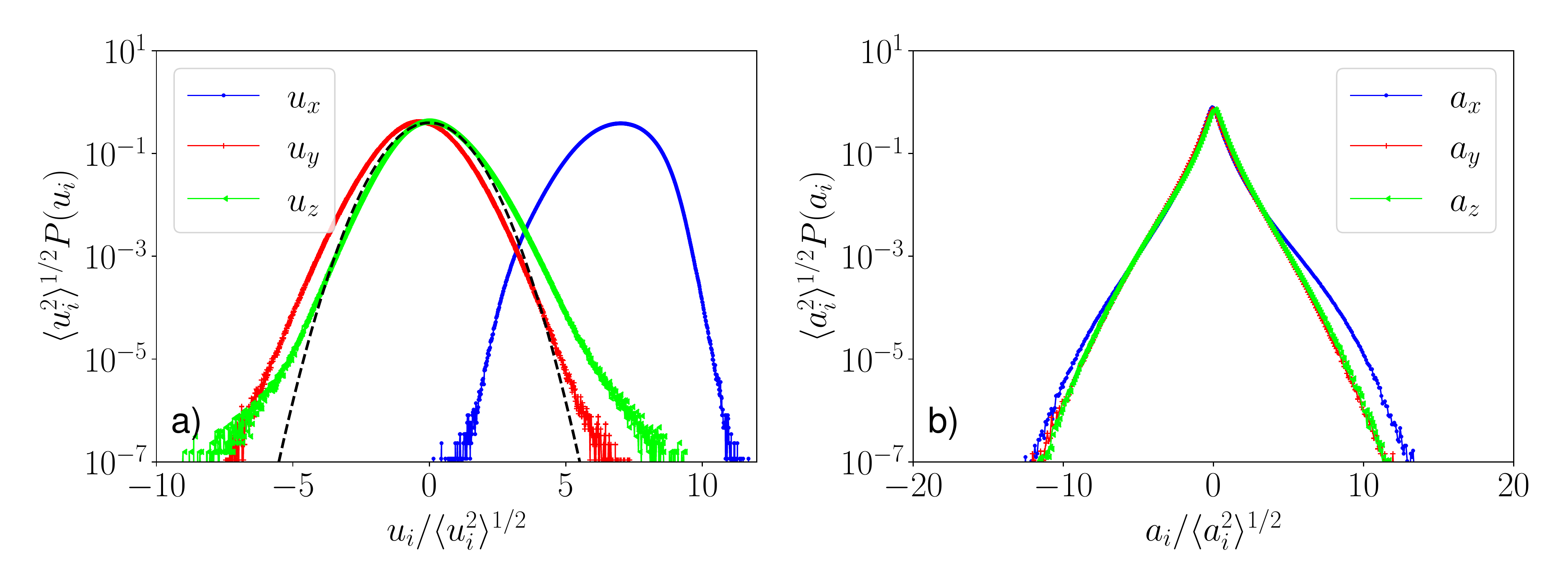}
    \caption{Standardised PDFs of velocity a) and acceleration b) components in the logarithmic region of a ZPG turbulent boundary layer over a flat plate using the root-mean-square (rms) of velocity fluctuations for the velocity 
    and the rms of the acceleration for acceleration PDFs. The dashed line in a) corresponds to a Gaussian with zero mean and unit variance. \moritz{updated}}
    \label{fig:BLua}
\end{figure}

PDFs of velocity and acceleration components in stream-wise, span-wise and wall-normal directions are shown in fig.~\ref{fig:BLua} a), b), respectively. As can be seen from in subfigure a), the PDF of the stream-wise velocity component $u_x$ has nonzero mean and clearly sub-Gaussian tails \moritz{with a flatness value of 2.69}. The PDFs of the
velocity in span-wise $y$-direction and wall-normal $z$-direction both have \moritz{super}-Gaussian tails but approximately zero mean \moritz{and flatness values of 3.29 and 3.36, respectively}. 
These clear indications of anisotropy are not present in the PDFs of stream-wise, span-wise and wall-normal acceleration components, which fluctuate very similarly as can be seen from the data shown in subfigure b).
\moritz{However, the boundary layer dataset has less extreme acceleration events compared to the other datasets, in this context we point out that Taylor-scale Reynolds number of the the turbulent ZPG boundary layer is smaller in comparison with the other datasets, see table~\ref{tab:flow-properties}.
}
For a further analysis of Eulerian and Lagrangian statistics obtained from the same dataset, see Ref.~\cite{Bross2023}.

Having discussed velocity and acceleration statistics, we now focus on -- to the best of our knowledge first -- measurements of curvature and torsion fluctuations in the ZPG boundary layer.  
The most striking observation here is that the right tail of the curvature PDF differs significantly from the curvature PDFs of the previously discussed datasets and from those reported in Refs.~\cite{Xu,Braun,Alards2017}. The left tail of the curvature PDF is still linear, however, low curvature events are more likely in the turbulent boundary layer compared to the aforementioned datasets.
The right tail of the PDF does not have power law form, and is much lighter than for the for the aforementioned datasets. 
To explain this observation, we recall that the
formula for the curvature is $\kappa = a_n/u^2$, and extreme curvature events are generated mainly by low-velocities events, rather than by high acceleration events \cite{Xu}. 
A strong unidirectional flow results in low-velocity events to be less likely, resulting in less high curvature events.
While the right tail of the curvature PDF is strongly influenced by large-scale motions, the torsion PDF has power law tails with the same exponents as observed for the other datasets discussed here. 
Importantly, high torsion values are less likely and low-torsion events more likely to occur 
in the ZPG turbulent boundary layer as compared to von Kármán flow and RBC, which can be
interpreted as trajectories in a flow with a strong stream-wise velocity component to be less twisted. 

\begin{figure}
    \centering
    \includegraphics[width = \textwidth]{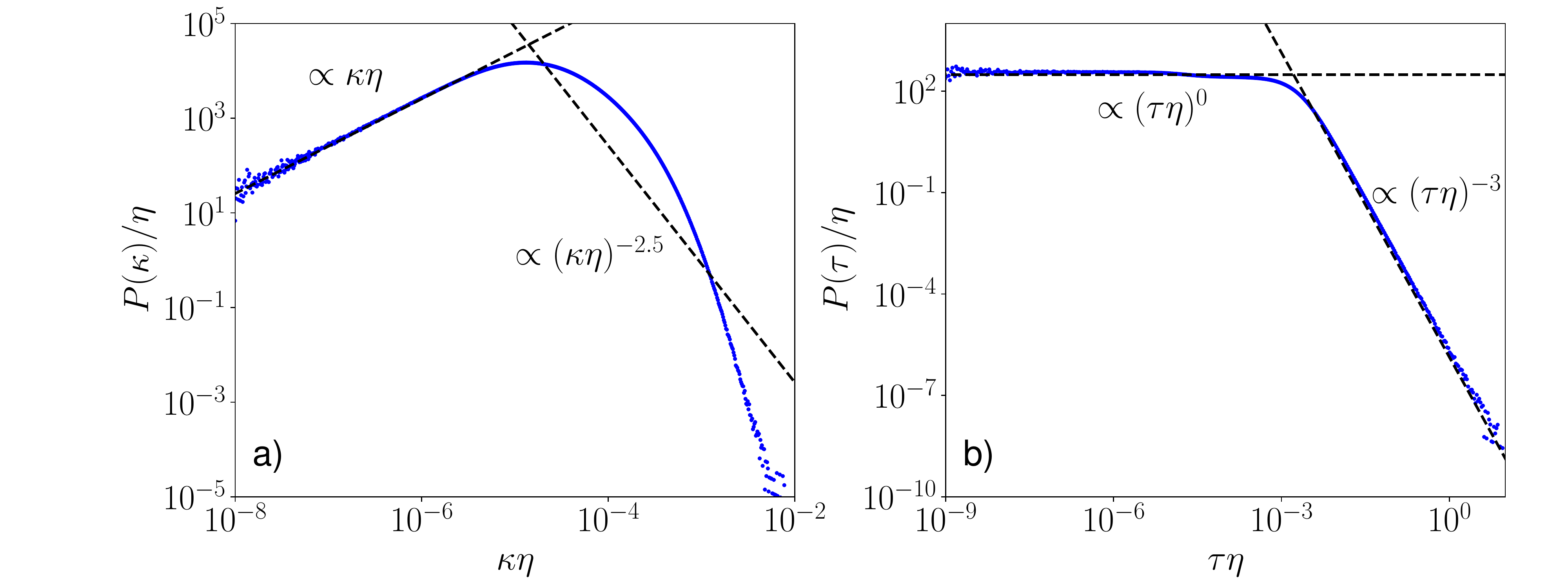}
    \caption{a) Curvature and b) torsion PDFs for the logarithmic layer in a ZPG turbulent boundary layer. Deviations of the expected power laws for the curvature are discussed in the text. \moritz{updated}}
    \label{fig:BLcurv_tors}
\end{figure}

\subsection{Quantitative comparison of curvature and torsion statistics}
Curvature and torsion have units of inverse length and can be made dimensionless by the ratio of the rms-value of the acceleration and the velocity variance \cite{Xu}, that is, 
\begin{align}
\label{eq:kappa-scaling}
    \kappa &= \frac{|\bm{a}_n|}{|\bm{u}|^2} \propto \frac{\langle a^2 \rangle^{1/2}}{\langle u^2\rangle} \ , \\
    \label{eq:tau-scaling}
    \tau &= 
    \frac{\bm{u} \cdot (\bm{a} \times \dot{\bm{a}})
    }{
    |\bm{u}|^3\kappa^2} 
    \propto 
    \frac{\langle u^2 \rangle^{1/2} \langle a^2\rangle^{1/2}
    \frac{\langle u^2 \rangle^{1/2}}{\langle a^2 \rangle^{1/2}}}
    {\langle u^2\rangle^{3/2}
    \left(\frac{\langle a^2 \rangle^{1/2}}{\langle u^2\rangle}\right)^{2}}
    =
    \frac{\langle a^2 \rangle^{1/2}}{\langle u^2\rangle}.
\end{align}
Based on Kolmogorov's original scaling arguments \cite{Kolmogorov41a,Kolmogorov41b}, Heisenberg and Yaglom \cite{Heisenberg48,Yaglom49} derived the following scaling relation for the acceleration covariance (cf. \cite{Porta_nature,Voth2002})
\begin{equation}
    \langle a_i a_j\rangle = a_0 \varepsilon^{3/2} \nu^{-1/2} \delta_{ij} \ ,
\end{equation}
where $a_0$ does not depend on the Reynolds number for K41 scaling. In conjunction with eqs.~\eqref{eq:kappa-scaling} and \eqref{eq:tau-scaling}, Heisenberg-Yaglom scaling results in a Reynolds-number-dependent non-dimensionalisation of curvature \cite{Xu} and torsion and thus enables a comparison of PDFs for data at different Reynolds numbers. For von K\'arm\'an flow, curvature PDFs for $ 200\leqslant Re_\lambda \leqslant 815$
indeed collapse onto a master curve \cite{Xu}.

\begin{figure}
    \centering
    \includegraphics[width =\textwidth]{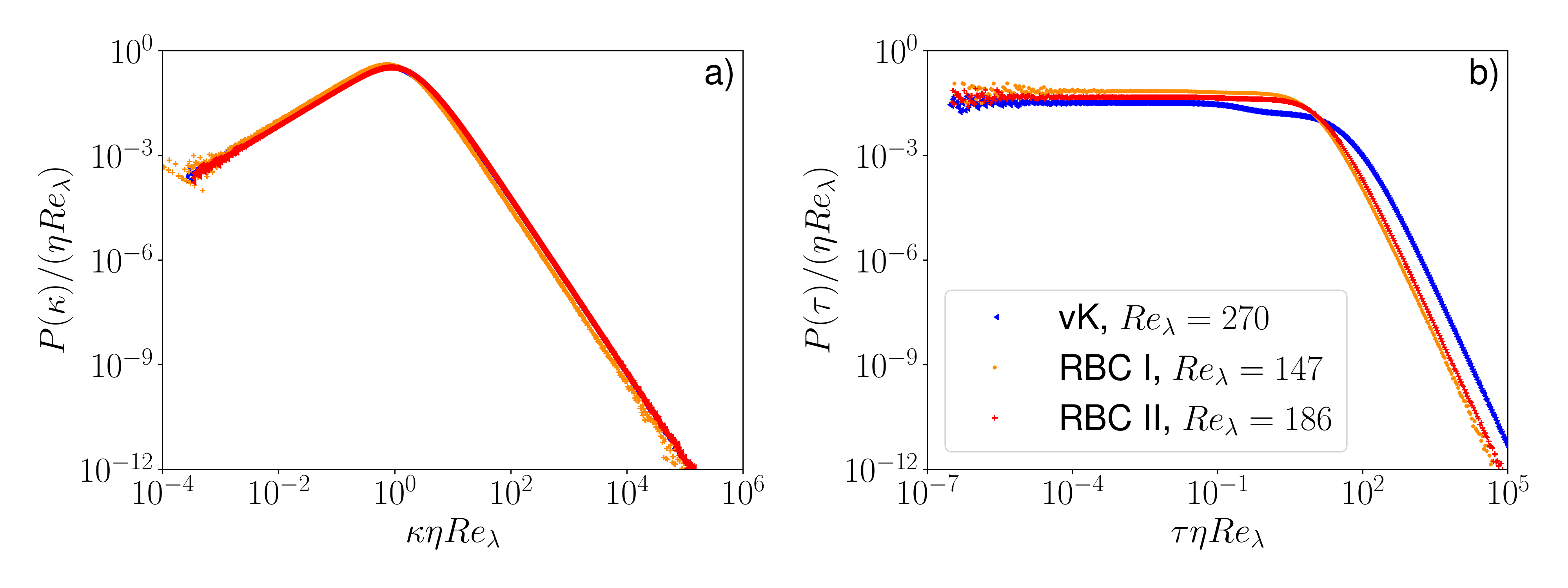}
    \caption{Heisenberg-Yaglom scaled
	    PDFs of a) curvature and b) torsion for von K\'arm\'an flow and Rayleigh-B\'enard convection, for the latter in the bulk only. \moritz{updated}
          }
	\label{fig:curvtors_comp}
\end{figure}

Here, we use the same ansatz to compare the PDFs of curvature and torsion for von Kármán flow and Rayleigh-Bénard convection at different Reynolds numbers with results shown in figs.~\ref{fig:curvtors_comp} a), b). 
\moritz{ The curvature PDFs
collapse onto master curves, while the torsion PDFs become close after rescaling but do not collapse on a master curve. A residual and consistent $Re_\lambda$-dependence can be observed in the data presented in fig.~\ref{fig:curvtors_comp} b), with high torsion events becoming more likely with increasing $Re_\lambda$. There may be several reasons for this. As torsion measurements probe smaller scales, the measurements may be more sensitive to intermittency, that is, a $Re_\lambda$-dependence of $a_0$. However, the torsion measurements need to be taken with caution here, as outlined in section \ref{sec:methods}. 
The collapse on master curves for the curvature PDFs and approximately so for the torsion PDFs} confirms that in the bulk, the form of the curvature and torsion PDFs are only determined velocity and acceleration fluctuations
and do not depend on the geometry of the flow or the type of turbulence production, as suggested in Ref.~\cite{Alards2017}, at least for RBC or, more generally, relatively low levels of anisotropy. 
As the curvature PDF for the turbulent boundary layer is phenomenologically different from the curvature PDFs of the von K\'arm\'an  and RBC data, is it is not included in the comparison. However, even the torsion PDF or the turbulent boundary layer could not be re-scaled with $Re_\lambda$ using Heisenberg-Yaglom scaling to fit on the torsion master curve,  
which may be connected to the low value of Taylor-scale Reynolds number, \moritz{$Re_\lambda = 108$}.

\section{Curvature vector statistics }
\label{sec:curv_vector}
As seen in previous sections, large-scale turbulent fluctuations in all presented datasets are statistically anisotropic, either due to the motion and location of the propellers in the von K\'arm\'an experiment, the direction of a temperature gradient and the ensuing presence of the LSC in Rayleigh-B\'enard convection (i.e. the remnants of a superstructure) or 
a strong unidirectional flow and 
the presence of large-scale coherent 
structures in the ZPG turbulent boundary layer.

As discussed in section \ref{sec:results}, curvature fluctuations are insensitive to anisotropy,  unless the degree of anisotropy is appreciably large -- a quantification thereof may be addressed in future work -- and differences in curvature fluctuations between the respective datasets are only due to Reynolds number. 
This may not be too surprising, since curvature as a global observable is coordinate-independent and mixes information pertaining to the different spatial directions. As such, it can only provide information on how curved trajectories in general are in a flow. To provide further insight into how large-scale motion affects the geometry of particle trajectories, we now focus on 
the statistics of the curvature vector, which can give a measure of anisotropy of the flow as detailed in sec. \ref{sec:geometry}. 

\begin{figure}
    \centering
    \includegraphics[width = \textwidth]{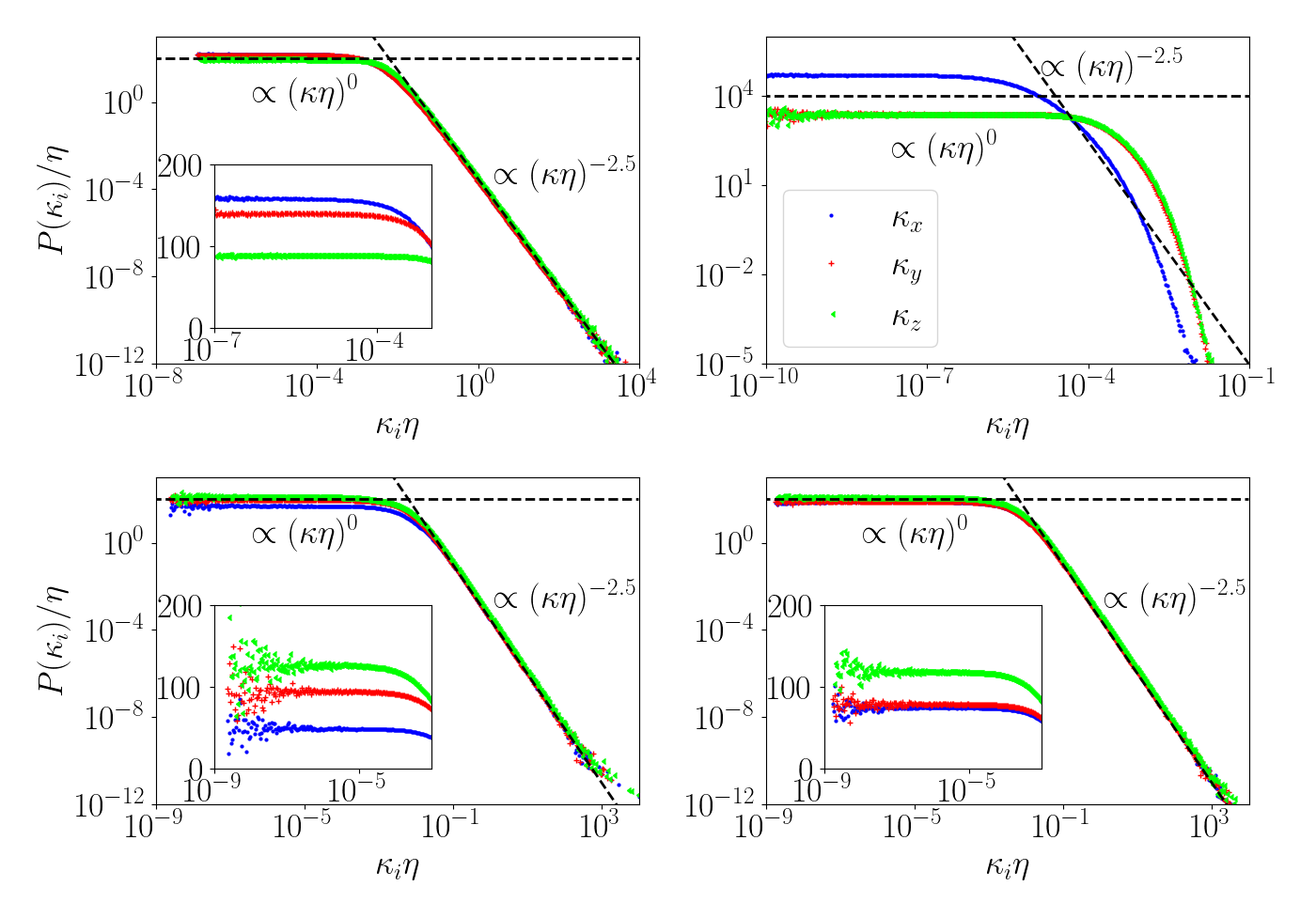}
    \caption{PDFs of the non-dimensionalised components of the curvature vector for different datasets in log-log representation, a) von Kármán flow b) Boundary layer, c) RBC I ($Ra= 5.3 \cdot 10^8$), d) RBC II ($Ra= 1.53\cdot 10^9$). References lines $(\kappa \eta)^0$ and $(\kappa \eta)^{-2.5}$ are shown for all cases. In a), c), d) insets show PDFs in logarithmic-linear representation.\moritz{updated}}
    \label{fig:curvvec}
\end{figure}

Figure \ref{fig:curvvec} shows the PDFs of the absolute value of the projection of the curvature vector onto the $x$-, $y$- and $z$-directions for the four different datasets. 
A few general observations can be made for all datasets. Firstly, the left tails of the PDFs are constant in all cases. The corresponding plateaux distinguish between the different directions, as can be seen from the insets in figs.~\ref{fig:curvvec} a), c), d) and directly in fig.~\ref{fig:curvvec} b). Second, the right tails, which describe high-curvature events in the respective spatial directions have the same power law as the curvature PDF for von K\'arm\'an 
flow and Rayleigh-B\'enard convection, which may be expected as discussed in sec.~\ref{sec:geometry}. For the ZPG boundary layer, the high-curvature tails 
again do not follow a power law and are qualitatively similar to the full curvature PDF. 

\begin{figure}
    \centering
    \includegraphics[width = \textwidth]{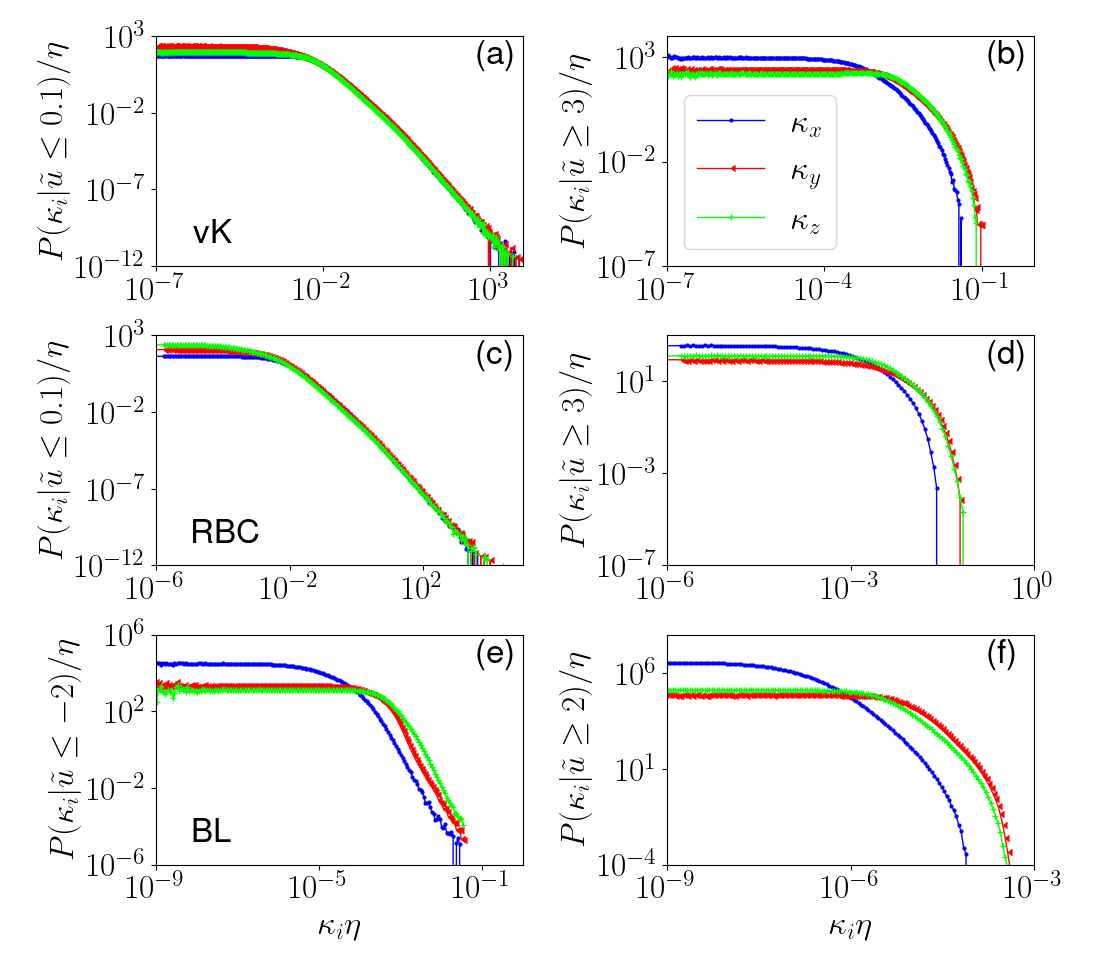}
    \caption{\moritz{PDFs of the curvature vector components, conditioned on the velocity $\tilde{u}$. For vK and RBC $\tilde{u} = u_x / \sigma_{u_x}$, for BL $\tilde{u} = \frac{u_x-\overline{u_x}}{\sigma_{u_x}}$. Conditioning on high velocities suppresses high curvature events in all cases.}}
    \label{fig:curvvec_cond}
\end{figure}

To give a plausible explanation for the differences in likelihood of low-curvature events in the different directions, we connect large-scale, i.e. velocity, statistics with curvature vector statistics for and across all datasets.
\moritz{
For that, we calculate the PDFs of the curvature vector components, 
conditioned on small velocities ($|u_i/\sigma_{u_i}| \leq 0.1$) or large velocities ($|u_i/\sigma_{u_i}| \geq 3$) of one component.
These values are chosen for all datasets and all components, expect for the stream-wise direction of the boundary layer, where the PDFs are conditioned on $\frac{u_x-\overline{u_x}}{\sigma_{u_x}}\leq -2$ 
or $\frac{u_x-\overline{u_x}}{\sigma_{u_x}}\geq 2$ for low and high velocity events, respectively. 
For simplicity, only the PDFs conditioned on $u_x$ are shown, the remaining PDFs show the same effects. An intuitive expectation would be that trajectories are less curved in a direction if the velocity in this direction is high. This would result in a lower probability of high curvature events compared to the other directions (right tail), which indeed can be seen in the right column of fig.~\ref{fig:curvvec_cond}. It can also be seen that high velocity events suppress high curvature events, resulting in a change of the power law for high curvature values, similar to the shape of the curvature PDF of the turbulent ZPG boundary layer. 
On the other hand, having a small velocity in one direction (fig.~\ref{fig:curvvec_cond} left column) leads to an decreased probability of low curvature events for the von Kármán dataset and the RBC. 
This is not true for the boundary layer. 
Due to the nature of this flow and the filtering approach we took, small velocity events have still velocities significantly larger than zero.
As a result low curvature events in $x$-direction are more likely than in $y$- and $z$-direction. What can also be noted here, that if high velocity events are filtered, the right tail of the PDFs flatten and  follow a power law. This shows that the effect of a non-zero mean dominates the statistics of the curvature vector and is another indicator that the changes of the power law for the right curvature PDF tail of the turbulent ZPG boundary layer origins in a strong mean flow.} 

\moritz{In a geometric  sense, this implies that trajectories are rarely curved against the direction of the mean flow. They predominantly either meander, that is, they bend in spanwise direction or have helical shape with rotation axis aligned with the mean flow direction.}

Comparing all datasets, we point out that the differences between the low-curvature plateaux in the PDFs for von K\'arm\'an flow and bulk Rayleigh-B\'enard convection shown in the insets of figs.~\ref{fig:curvvec} a), c) and d), respectively, are all of the same order of magnitude. Interestingly, according to the 
curvature vector statistics von K\'arm\'an flow is geometrically less isotropic than Rayleigh-B\'enard convection in the bulk. For the latter, anisotropy -- in the sense of differences between the statistics of the curvature vector components -- decreases with increasing $Ra$, as can be expected.

Generally speaking, strong velocity events in one direction lead to an increase of low curvature events and a decrease of high-curvature events in that same direction. Based on the data of the turbulent ZPG boundary layer, we expect that a mean flow in one direction suppresses effects of extreme events on the curvature vector, especially for its component in stream-wise direction.  
This poses the question as to the effect of extreme acceleration events. It is conceivable that more extreme fluctuations at small scale in one direction leads to an increased probability of high curvature events in that direction. This would be physically intuitive as extreme events indicate turbulence, hence vortices, where trajectories are expected to be more curved, but this remains to be investigated.  

\moritz{
Before concluding, we briefly compare the aformentioned results with standard methods of anisotropy detection. 
Anisotropy in the Lagrangian frame of reference is mostly studied by comparison of Lagrangian structure functions and spectra in different reference directions. For von K\'{a}rm\'{a}n flow, such measurements reveal differences in the scaling constants of both velocity and acceleration prefactors \cite{LaPorta2001,Voth2002,Ouellette2006,Huck2017,Huck2019}, which implies that the large-scale anisotropy that is present in the flow due to large-scale strain \cite{Huck2017} affects turbulent fluctuations at the small scales. Such measurements provide global statistical information on the level of anisotropy on large and small scales, they do not encode information on the type of anisotropy. 
More detailed information can be obtained in the Eulerian frame of reference through the Lumley triangle \cite{Schumann1977,Lumley1977,Pope2000}, based on measurements of the second and third invariant of the Reynolds stress tensor. The Lumley triangle distinguishes between types of anisotropy, such as disk-like and rod-like flow structures. For von K\'{a}rm\'{a}n flow at ${\rm Re}_\lambda = 815$ flow structures are predominantly disk-like, and occasionally rod-like, \cite{Risius2015} (see Fig.~13). 
For slowly rotating RBC, flow structures in the bulk appear to be mostly rod-like \cite{Kunnen2010} in the direction of the temperature gradient (CHECK), which has been attributed to the influence of the LSC. 
For a canonical channel flow, flow structures in the log-law region are mainly rod-like \cite{Kim1987,Pope2000}, with the principal axis preferentially aligned with the mean flow direction (CHECK). 
The measurements of the curvature vector statistics are commensurate with the presence of such structures. For RBC and the turbulent ZPG-BL, low-curvature events are more likely in the direction of either the temperature gradient ($z$-direction) or the mean flow ($x$-direction), respectively.  In the case of von K\'{a}rm\'{a}n flow, low-curvature events are less likely in the contractile ($z$-direction) of the flow, where the flow velocity is lower and more likely in the extensional directions where the flow velocities are higher. In summary, the comparison with standard Eulerian measurements supports the hypothesis that trajectories are less likely to bend in the direction of strong flow.
}

\section{Conclusions} \label{sec:conclusions}

Here, we compared Lagrangian statistics for four experimental datasets of three different types of
turbulent flows, focusing on the effect large-scale motion on the geometry of tracer particle trajectories. To observe and quantify the latter, we introduced the curvature vector and calculated statistics of its projections in the spatial directions determined by the experimental apparatus.  

The datasets we considered were von Kármán flow, Rayleigh-B\'enard convection in the bulk at two different Rayleigh numbers
and a turbulent zero-pressure-gradient boundary layer in the logarithmic region. 
The Taylor-scale Reynolds numbers are $Re_{\lambda}
= 270$ for the von Kármán flow, \moritz{$Re_{\lambda} = 147$ and $186$} for the RBC with
$Ra = 5.3 \cdot 10^8$ and $Ra = 1.53 \cdot 10^9$ respectively and $Re_\tau = 2295$ $( \moritz{Re_{\lambda}
= 108)}$ for the turbulent boundary layer. To establish the baseline statistical characteristics of the flows, we first calculated PDFs of instantaneous
velocity, acceleration, curvature and torsion of Lagrangian trajectories. 
Rayleigh-B\'enard convection (bulk only) and von K\'arm\'an data behave statistically 
very similar: the PDFs of the velocity components are approximately Gaussian, acceleration PDFs have the expected wide tails.
The power laws found for the tails of the curvature PDFs agree with previous
results from numerical simulations of homogeneous and
isotropic turbulence (HIT) \cite{Braun}, experiments of von Kármán flow
\cite{Xu} and numerical simulations of Rayleigh-B\'enard convection \cite{Alards2017}. 
Similarly, the torsion PDFs in all three datasets show the same
scaling as for numerical simulations of homogeneous isotropic turbulence \cite{Scagliarini} and RBC \cite{Alards2017}. 
Both curvature and torsion PDFs can be re-scaled using Heisenberg-Yaglom scaling to adjust for different Reynolds numbers and collapse onto a master curve. 
That is, the geometry and flow type
do not seem to have an influence of any of these statistics, as long as there
is no mean flow. The turbulent boundary layer has a mean flow in $x$-direction. The PDFs of all velocity components are \moritz{near}-Gaussian, with approximately zero mean on wall-normal and span-wise directions. The acceleration PDFs do not seem to be influenced by that, we found
PDFs with wide tails, which, when normalised by the respective standard deviations, collapsed into
one. The curvature PDF has the same tail for small values of the curvature but
deviates from the -2.5 power law for high values. This is most likely due to 
the strong unidirectional flow suppressing high-curvature events and a strong unidirectional flow breaks the apparent universal form of the curvature PDF. 
The torsion PDF 
is unaffected by the geometry and the same power laws tails as for homogeneous isotropic turbulence are found.

For the components of the curvature vector, however, we observe marked differences between datasets and spatial directions.  Firstly, the projection of the curvature vector is much more likely to be smaller in stream-wise direction compared to the span-wise and wall-normal direction for the ZPG boundary layer, reflecting the physical intuition that trajectories of particles with strong stream-wise velocity are less curved against the flow and more likely to meander in wall-normal and span-wise directions. Secondly, also for von K\'am\'an flow and Rayleigh-Benard convection we find differences in the low-curvature tails of the curvature vector PDFs in the respective spatial directions. A comparison between velocity statistics and curvature vector statistics reveals that low-curvature events occur mostly in directions where velocity fluctuations are stronger. This is commensurate with the observations made for the turbulent boundary layer.

In summary, through connecting the statistics of the curvature vector with that of velocity fluctuations we demonstrate that large-scale motion in a given spatial direction results in meandering rather than helical trajectories.  For the turbulent boundary layer, this is commensurate with the current understanding of superstructures \cite{Hutchins2007a,Kevin2019}. 
However, further analysis is required to distinguish between  trajectories within a superstructure and the background. This requires the calculation of curvature statistics conditioned on the presence of a large-scale coherent structure, which in turn requires a clear identification thereof. For RBC, turbulent superstructures can be found by data-driven means \cite{Schneide2018, Vieweg2021}, hence RBC lends itself well for a first investigation and quantification of the effects of turbulent superstructures on the geometry of tracer particle trajectories.   
Further work should also include a quantification of the observations made here through e.g. the calculation of joint statistics of \moritz{acceleration} and curvature vector components, and specifically for RBC the detection of potential correlations with between high-curvature events and ejecting plumes \cite{Schumacher2009}. The latter requires statistics conditioned on temperature.  
Finally, regarding the connection between Eulerian and Lagrangian statistics, one may try to connect curvature vector statistics with vortical structures. We will address these questions in future projects. 

\section*{Acknowledgements}
This project was inspired by Bruno Eckhardt's interest in curvature and torsion of tracer particle trajectories. Sadly, he passed away on August 7th 2019, and we do not know which research direction he would have pursued.
We thank Michael Wilczek and Lukas Bentkamp for helpful discussions \moritz{and Philipp Godbersen for the calculation of the Kolmogorov microscales for the ZPG turbulent boundary layer}. 
Computational resources on Cirrus  ({\tt www.cirrus.ac.uk}) have been obtained through Scottish Academic Access. 
This work received funding from Priority Programme SPP 1881 ``Turbulent Superstructures" of the Deutsche Forschungsgemeinschaft (DFG, grant numbers LI3694/1, KA1808/21, BO5544/1 and SCHR1165/5).
Yasmin Hengster was also supported by \moritz{the School of Mathematics at the University of Edinburgh}.
We acknowledge funding from the European High-Performance Infrastructures in Turbulence
(EuHIT) consortium for the DTrack measurement campaign at the von Kármán flow facility GTF3 and the support of the staff at MPIDS in G{\"o}ttingen, in particular Eberhard Bodenschatz.

\bibliographystyle{unsrtnat}
\bibliography{bib}

\end{document}